\documentclass[letterpaper,journal]{IEEEtran}
\usepackage{amsmath}
\usepackage{array}
\usepackage{tabularx}
\usepackage[caption=false,font=footnotesize,labelfont=rm,textfont=rm]{subfig}
\usepackage{graphicx}
\usepackage{tikz}
\usepackage{orcidlink}
\usepackage{makecell}
\usepackage[ruled,linesnumbered]{algorithm2e}
\usepackage{booktabs}
\usepackage{amssymb}
\usepackage{cite}
\usepackage{threeparttable}
\usepackage{microtype}
\usepackage{hyperref}
\usepackage{xcolor}
\hypersetup{colorlinks=true,linkcolor=blue,citecolor=blue,urlcolor=blue}

\newcommand{\Sat}{\mathcal{V}}
\newcommand{\Adm}{\mathcal{A}}

\begin{document}

\title{Relieving Hotspot Congestion in Large-Scale Low Earth Orbit Satellite Networks via Distributed Dual-Timescale Traffic Splitting}

\author{
Shuyang~Li\orcidlink{0000-0001-5365-4137},~\IEEEmembership{Member,~IEEE,}
Guanchen~Wu\orcidlink{0009-0000-3349-472X},
Qiang~Wu\orcidlink{0000-0002-9467-1491},~\IEEEmembership{Member,~IEEE,}
Ran~Wang\orcidlink{0000-0001-5601-0513},~\IEEEmembership{Senior~Member,~IEEE,}\\
Jie~Hao\orcidlink{0000-0002-1269-2097},~\IEEEmembership{Member,~IEEE,}
Chao~Ren\orcidlink{0000-0001-9096-8792},~\IEEEmembership{Member,~IEEE,}
and~Hongke~Zhang\orcidlink{0000-0001-8906-813X},~\IEEEmembership{Fellow,~IEEE}
\thanks{This work has been submitted to the IEEE for possible publication. Copyright may be transferred without notice, after which this version may no longer be accessible.}
\thanks{This work was supported in part by the National Science and Technology Major Project under Grant 2026ZD1307100, in part by the Zhejiang Provincial Pioneer Program under Grant 2026SDXT011, in part by the Major Scientific and Technological Special Project of Jiangsu Province under Grant BG2025003, and in part by the Frontier Technologies Research and Development Program of Jiangsu under Grant BF2024067. Shuyang Li and Guanchen Wu contributed equally to this work. \textit{(Corresponding authors: Qiang Wu; Ran Wang.)}}
\thanks{Shuyang Li, Ran Wang, and Jie Hao are with the College of Computer Science and Technology, Nanjing University of Aeronautics and Astronautics, Nanjing 211106, China (email: shuyang.li@nuaa.edu.cn; wangran@nuaa.edu.cn; haojie@nuaa.edu.cn). Ran Wang is also with the College of Computer and Data Science, Fuzhou University, Fuzhou 350108, China.}
\thanks{Guanchen Wu is with the School of Mathematics, Jilin University, Changchun 130012, China (email: wugc1023@mails.jlu.edu.cn).}
\thanks{Qiang Wu is with the College of Computer Science and Technology, Zhejiang University, Hangzhou 310027, China (email: wu.qiang@zju.edu.cn).}
\thanks{Chao Ren is with the Aerospace Information Research Institute and the National Key Laboratory of Target Cognition and Application Technology, Chinese Academy of Sciences, Beijing 100094, China (email: renchao@aircas.ac.cn).}
\thanks{Hongke Zhang is with the School of Electronic and Information Engineering, Beijing Jiaotong University, Beijing 100044, China (email: hkzhang@bjtu.edu.cn).}}

\maketitle

\begin{abstract}
Sixth-generation mobile systems envision ubiquitous connectivity through non-terrestrial networks, in which low Earth orbit (LEO) satellite networks play a key role by extending broadband access beyond terrestrial infrastructure.
Traffic hotspots in LEO satellite networks congest inter-satellite links, and short flows leave end-to-end congestion control little time to react.
In distributed traffic splitting, recomputing splits at every queue change increases onboard computation, whereas periodic splits miss such changes and withhold zero-weight ports from faster queue-based reweighting.
We formulate the distributed slow-layer traffic-splitting problem, whose objective combines throughput, queueing delay, traffic loss, and short-flow completion time, and propose Balanced Routing with Admissible Ports and Congestion Control at Two Timescales (BRACE).
Its slow layer assigns destination-aware splits with a positive weight on every admissible port, and its fast layer reweights them in constant time per hop using instantaneous local queues.
In 144-satellite flow-level simulations, BRACE achieves the highest reward among the tested distributed methods in all seven hotspot conditions and 28--45\% lower drop rates than the lowest-loss baseline, a tuned congestion-aware heuristic, in the three highest-load conditions.
Without retuning, its drop rates are 26--76\% lower than that baseline's at calibrated loads in constellations of up to 1584 satellites.
\end{abstract}


\section{Introduction}
Wide-area coverage and connectivity in underserved regions are central objectives of sixth-generation mobile systems.
The International Mobile Telecommunications framework for 2030 and beyond identifies ubiquitous connectivity and interworking with non-terrestrial networks~\cite{itu2023imt2030}.
Among these networks, low Earth orbit (LEO) satellite networks can extend broadband access beyond terrestrial infrastructure~\cite{giordani2021ntn}.
With inter-satellite links (ISLs), such a network relays traffic between distant ground endpoints over multiple satellites, so an ISL can carry traffic from many ground regions.
Efficient transport across ISLs is therefore essential, yet path structure, delay, and link utilization change as satellites move~\cite{kassing2020exploring}.

Geographical traffic imbalance compounds these dynamics.
ISLs serving high-demand regions may become congested while capacity elsewhere remains underused~\cite{taleb2009elb,li2026tpbr}.
Relieving this congestion requires both distributing traffic across available paths and responding promptly to queue buildup.
End-to-end congestion control, however, changes sending rates rather than paths, and does so only after feedback traverses the network.
Because flow sizes are heavy-tailed, most flows are short~\cite{crovella1997self} and can complete within a few round-trip times (RTTs), leaving such feedback little time to help them.
These limitations motivate congestion control inside the satellite network, where traffic can be redirected as queues form.

Designing such control involves a trade-off between responsiveness and overhead.
Decisions based on period-averaged measurements can miss queue changes between updates.
Frequent route recomputation or policy inference increases onboard computation, while collecting fresh network-wide demand adds signaling overhead~\cite{guo2026test}.
The resulting challenge is to redistribute traffic promptly using local information while keeping computation and signaling overhead manageable.

Existing approaches meet this challenge only in part.
Local reactive routing schemes redistribute traffic in response to congestion notifications or queue information rather than through periodically planned traffic fractions~\cite{taleb2009elb,deng2023dbpr}.
Hybrid global-local (HGL) routing combines planned traffic distribution with local adjustments, but relies on network-wide demand prediction~\cite{liu2017hgl}.
In-network load balancers use congestion feedback or local queues for fast path or next-hop selection within candidates supplied by routing~\cite{alizadeh2014conga,ghorbani2017drill}; in a LEO network, which candidates to offer and how strongly to prefer each must themselves be decided per destination as the topology changes.

These approaches leave a coordination gap for distributed LEO control: how to combine destination-aware traffic distribution and prompt congestion responses using locally available information as the topology changes.
A natural combination lets each satellite periodically plan traffic splits from local information and reweight them using a faster queue-based rule.
The two timescales are coupled, however: if the periodic split assigns zero weight to a port, fast reweighting of that split cannot use the port for congestion relief, whereas keeping every port available admits detours that lengthen paths.
The key questions are therefore which forwarding alternatives should remain available for congestion relief without excessive detours, and how fresh the local queue information must be for the faster rule to help.
The main contributions are as follows.
\begin{enumerate}
\item We formulate the \emph{distributed slow-layer traffic-splitting problem} (P1).
In P1, each satellite uses local information to set split weights over \emph{admissible ports} for each reachable nonlocal destination once per control period. Faster local congestion responses follow fixed analytic rules.
The objective accounts for throughput, queueing delay, traffic loss, and the flow completion time (FCT) of short flows.

\item We design \emph{Balanced Routing with Admissible Ports and Congestion Control at Two Timescales} (BRACE) as a \emph{distributed dual-timescale framework} with an explicit interface between routing plans and local congestion responses.
Once per control period, the slow layer assigns destination-aware soft splits over locally admissible ports and keeps a positive weight on every admissible port.
At each path-selection event, the fast layer reweights the split in constant time per hop using instantaneous local queues.

\item We conduct \emph{comparative flow-level experiments} under orbital dynamics at real speed and under a stress setting with accelerated motion and random link failures.
With validation-selected parameters, BRACE achieves the highest mean reward among the tested distributed methods in all seven hotspot conditions and a lower drop rate than the lowest-loss distributed baseline in the three highest-load conditions.
The high-load ablations also show that queue snapshots refreshed every 20\,ms perform close to instantaneous queues, whereas refreshing them every 500\,ms yields lower reward than running without the fast layer.
The scalability experiments extend to 1584 satellites at calibrated loads without retuning BRACE's control parameters.
\end{enumerate}

Section~\ref{sec:related} reviews satellite routing, congestion control, and in-network load balancing.
Section~\ref{sec:system} describes the LEO network model, flow-level data plane, and performance metrics.
Section~\ref{sec:problem} formulates P1, defining the control timescales, admissible ports, and network objective.
Section~\ref{sec:method} presents the fast and slow layers of BRACE, their interface, and computational complexity.
Section~\ref{sec:eval} evaluates BRACE against distributed baselines and a centralized reference, and examines timescale separation, scalability, robustness, and parameter sensitivity.
Section~\ref{sec:conclusion} summarizes the findings and outlines future research directions.

\section{Related Work}
\label{sec:related}
Table~\ref{tab:related} compares the mechanisms and properties of representative approaches; checkmarks indicate explicit support in the cited work.
\emph{Planned traffic fractions + local adjustment} denotes periodically planned split ratios realized or adjusted by a faster local mechanism.
\emph{Positive weight floor} denotes a strictly positive base weight on every admissible port.
\emph{Distributed decisions} are made at network nodes from locally available information, without centralized control or a network-wide demand matrix.

\begin{table*}[!t]
\centering
\caption{Comparison of coordination mechanisms and properties}
\label{tab:related}
\footnotesize
\setlength{\tabcolsep}{3pt}
\renewcommand{\arraystretch}{1.12}
\begin{threeparttable}
\begin{tabular}{@{}>{\raggedright\arraybackslash}p{5.71cm}c>{\centering\arraybackslash}p{1.9cm}>{\centering\arraybackslash}p{2.3cm}*{3}{>{\centering\arraybackslash}p{1.87cm}}@{}}
\toprule
& & \multicolumn{3}{c}{Mechanisms} & \multicolumn{2}{c}{Properties} \\
\cmidrule(lr){3-5}\cmidrule(l){6-7}
Work & Network & \makecell[t]{Flowlet\\forwarding} & \makecell[t]{Planned traffic\\fractions +\\local adjustment} & \makecell[t]{Positive\\weight floor} & \makecell[t]{Distributed\\decisions} & \makecell[t]{No\\training} \\
\midrule
ELB, TLR, DBPR, Feng et al. \cite{taleb2009elb,song2014tlr,deng2023dbpr,feng2024cooperative} & LEO &  &  &  & $\checkmark$ & $\checkmark$ \\
HGL~\cite{liu2017hgl} & LEO &  & $\checkmark$ &  &  & $\checkmark$ \\
LST~\cite{wu2025lst} & LEO &  &  &  &  & $\checkmark$ \\
MA-DRL, POMAP, Flow-centric \cite{lozano2025continual,li2025efficient,liu2024enabling} & LEO &  &  &  & $\checkmark$ &  \\
\midrule
CONGA, LetFlow, HULA, QALL \cite{alizadeh2014conga,vanini2017letflow,katta2016hula,liu2024qall} & Data center & $\checkmark$ &  &  & $\checkmark$ & $\checkmark$ \\
DRILL~\cite{ghorbani2017drill} & Data center &  &  &  & $\checkmark$ & $\checkmark$ \\
\midrule
TeXCP + FLARE \cite{kandula2005texcp,kandula2007dynamic} & Wide area & $\checkmark$ & $\checkmark$ & $\triangle$ & $\checkmark$ & $\checkmark$ \\
\midrule
\textbf{BRACE (ours)} & \textbf{LEO} & $\checkmark$ & $\checkmark$ & $\checkmark$ & $\checkmark$ & $\checkmark$ \\
\bottomrule
\end{tabular}
\begin{tablenotes}[flushleft]
\item[] $\triangle$ denotes partial support. TeXCP moves a small amount of traffic only to its least-utilized path.
\end{tablenotes}
\end{threeparttable}
\end{table*}

\subsection{Routing and Load Balancing in LEO Networks}
Load-balancing routing schemes for LEO networks move traffic away from congested ISLs using local congestion information.
In explicit load balancing (ELB)~\cite{taleb2009elb}, a satellite nearing congestion asks its neighbors to divert some traffic around it.
Traffic-light-based intelligent routing (TLR)~\cite{song2014tlr} switches between precomputed best and second-best routes according to traffic lights that indicate the congestion at the current and the next satellite.
Distance-based back-pressure routing (DBPR)~\cite{deng2023dbpr} weights links by a distance-based back-pressure metric and restricts forwarding to the rectangle between source and destination.
HGL routing~\cite{liu2017hgl} computes traffic splits from predicted network-wide demands through multicommodity flow (MCF) optimization and uses local congestion notifications to trigger adjustments.
Feng et al.~\cite{feng2024cooperative} combine minimum-hop constraints with local sending queues and next-hop load for distributed satellite--terrestrial routing.
The long-short-term (LST) routing algorithm~\cite{wu2025lst} couples long-term path planning with short-term transmit-power control based on channel and queue states.
ELB, TLR, DBPR, and Feng et al. use a single reactive routing mechanism.

Multi-agent deep reinforcement learning (MA-DRL) routing~\cite{lozano2025continual} selects next hops per packet with a shared deep Q-network (DQN) and continues to learn online.
Lyu et al.~\cite{lyu2024dynamic} incorporate energy and loss constraints into multi-agent reinforcement learning through Lagrange multipliers.
Our Pareto-optimal multi-agent proximal policy optimization packet routing framework (POMAP)~\cite{li2025efficient} models each satellite as a queue with active queue management and trains policies for Pareto-efficient packet routing.
We also combined DRL-based multipath route selection with a Stackelberg game that distributes traffic across the selected paths~\cite{li2025multipath}.
Because a forward pass per packet is costly, the flow-centric approach of~\cite{liu2024enabling} reuses routing decisions for packets in a flow and includes adaptive flow-routing updates.
These methods and other learning-based routing schemes~\cite{liu2026amg,lu2026lecmor} require training and neural-network inference.
In contrast, BRACE computes destination-aware splits from local information without demand prediction and reweights them with instantaneous local queues between updates, using analytic rules in both layers.

\subsection{End-Host and Network-Assisted Congestion Control}
Satellite-aware transport protocols exploit predictable dynamics to improve sender adaptation~\cite{li2026tidetcp,long2026multho}.
Cao and Zhang~\cite{cao2023satcp} use link-layer information to predict handovers and route updates and avoid unnecessary backoff in CUBIC, a congestion control algorithm for the Transmission Control Protocol (TCP).
SatPipe~\cite{zhao2025satpipe} adapts a congestion controller that estimates bottleneck bandwidth and round-trip propagation time, modifying its state machine to handle handovers and accelerate RTT estimation.
Yang et al.~\cite{yang2024maco} use mobility information to estimate the bandwidth--delay product for quick start and congestion window control in multipath QUIC.
Lai et al.~\cite{lai2025leocc} detect connection reconfigurations at the endpoint and regulate the sending rate with a reconfiguration-aware model of the network conditions.
Zhang et al.~\cite{zhang2022deepcc} use learning to adapt congestion control to application-specific performance objectives.
These end-host mechanisms can improve adaptation, including startup, but feedback delay still limits short-flow rate adjustment during unexpected ISL queue buildup.

Network-assisted control provides additional information. Data Center TCP (DCTCP)~\cite{alizadeh2010dctcp} uses explicit congestion notification (ECN) marks from switches to adjust sender windows.
Differential ECN marking~\cite{huang2025diffecn} protects flows that do not cause congestion, including newly arriving small flows.
The Explicit Control Protocol (XCP)~\cite{katabi2002xcp} and the Rate Control Protocol (RCP)~\cite{dukkipati2006rcp} compute explicit feedback or link rates inside routers.
These schemes act on sending rates and leave the paths of the flows unchanged, so they cannot move traffic from a congested ISL to underused ones.
BRACE retains end-host congestion control and redistributes traffic among admissible ISL ports using current local queues, without waiting for end-to-end feedback.

\subsection{In-Network Load Balancing}
\label{sec:in-network-lb}
Congestion aware balancing (CONGA)~\cite{alizadeh2014conga} uses congestion feedback between leaf switches to assign flowlets (bursts of packets within a flow) to paths.
LetFlow~\cite{vanini2017letflow} assigns flowlets to paths at random.
The hop-by-hop utilization-aware load balancing architecture (HULA)~\cite{katta2016hula} uses probes in programmable switches to place flowlets and handles topology changes and link failures.
Distributed randomized in-network localized load balancing (DRILL)~\cite{ghorbani2017drill} uses local queue occupancies at each switch to select a next hop for each packet from the candidates installed by routing.
Queue-behavior-aware localized load balancing (QALL)~\cite{liu2024qall} selects egress ports per packet or per flowlet, probabilistically from local queue behavior, including occupancy and queueing trends, without coordination among switches.
Proteus~\cite{hu2024proteus} combines RTT-level signals for initial path selection with sub-RTT cumulative sojourn time for re-routing around priority-flow-control pauses in lossless data centers.
These balancers choose among the paths or next hops that the data-center topology and its routing provide.
In the +Grid LEO topology considered here, however, each satellite has only four ISL ports, and the ports on shortest paths toward a destination change as inter-plane ISLs switch off at high latitudes.
Relieving a congested ISL may then require detour ports, whose use should be limited per destination.

Wide-area traffic engineering likewise separates path selection from online load adaptation.
In the Traffic Engineering Explicit Control Protocol (TeXCP)~\cite{kandula2005texcp}, each edge router shifts traffic among its paths using path-utilization feedback.
The flowlet aware routing engine (FLARE)~\cite{kandula2007dynamic} shifts traffic at flowlet boundaries to avoid packet reordering.
In semi-oblivious traffic engineering~\cite{kumar2018semi}, a centralized controller adapts sending rates over a static path set.
TeXCP also moves a small amount of traffic to its least-utilized path to reactivate unused paths.
For LEO constellations, Wu et al.~\cite{wu2025sate} compute network-wide traffic allocations at millisecond-scale latency with a graph neural network on a heterogeneous graph that follows the changing topology.
BRACE pairs local queue-based balancing with periodically updated, destination-aware limits on the ports each satellite may use, and keeps a positive weight on every admissible port so that each remains available to the fast layer.

\begin{figure*}[!t]
\centering
\includegraphics[width=\textwidth]{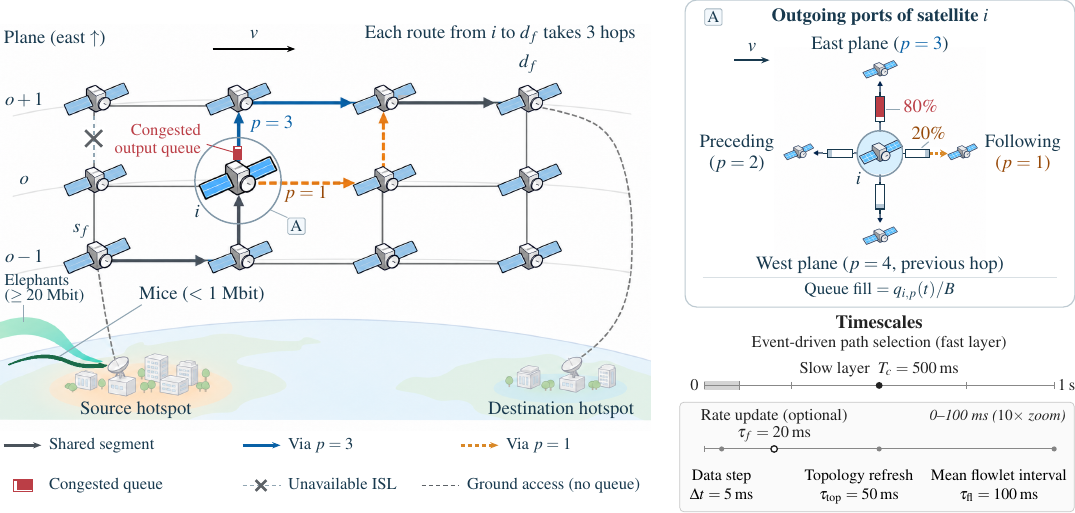}
\caption{Flow-level LEO network model and update intervals. Both routes from satellite $i$ to $d_f$ take three hops, and the route through $p=1$ avoids the congested output queue at $p=3$.}
\label{fig:system}
\end{figure*}

\section{System Model}
\label{sec:system}
We consider the inter-satellite segment of a LEO constellation that carries flows between ground users.
Fig.~\ref{fig:system} illustrates the network, outgoing-port queues, and update intervals.
The model is flow-level: every flow is represented by a sending rate, and every directed ISL by an aggregate fluid queue.
Packets, onboard processing, and channel errors are not modeled.
The data plane advances in steps of $\Delta t=5$\,ms, indexed by $t\in\{0,1,\dots\}$, with step $t$ starting at physical time $t\Delta t$, and the topology is refreshed every $\tau_{\mathrm{top}}=50$\,ms.
The control layers act through two quantities: forwarding weights $\varphi_{i,d,p}(t)\ge0$, by which satellite $i$ samples next ports toward destination $d$, and explicit rates $R_l(t)$ that directed ISLs may advertise to cap sending rates.
The slow layer updates the base port splits once per control period of $T_c=0.5$\,s, indexed by $k=\lfloor t\Delta t/T_c\rfloor$, with $T_c/\Delta t=100$ data steps per period.
When explicit-rate control is enabled, the advertised rates $R_l$ are refreshed every $\tau_f=20$\,ms.
An episode is one simulation run of $K$ control periods that starts with no active flows.
Sums over flows at step $t$ refer to the active flows after arrivals and before completion in that step.
Table~\ref{tab:notation} lists the main notation.

\begin{table}[!t]
\centering
\caption{Main notation}
\label{tab:notation}
\footnotesize
\begin{tabularx}{\columnwidth}{@{}l>{\raggedright\arraybackslash}X@{}}
\toprule
Symbol & Meaning \\
\midrule
$\Sat$, $N=|\Sat|$ & set and number of satellites \\
$p\in\mathcal P$, $n(i,p)$ & ISL port ($|\mathcal P|=4$); neighbor of $i$ behind $p$ \\
$l=(i,p)\in\mathcal L$ & directed ISL from $i$ through port $p$ \\
$e_l(t)$, $\mathcal G(t)$ & availability indicator of link $l$; graph of available links \\
$C_l$, $d_l(t)$ & link capacity (Mbps); propagation delay (s) \\
$q_l(t)$, $B$ & pre-update link queue in step $t$; buffer size (Mbit) \\
$\Delta t$, $t$ & data-plane step (5\,ms) and its index \\
$T_c$, $k$, $K$ & control period (0.5\,s); its index; periods per episode \\
$\tau_f$, $\tau_{\mathrm{top}}$ & explicit-rate and topology update periods \\
$\tau_{\mathrm{fl}}$ & mean interval between flowlet boundaries \\
$\lambda$ & flow arrival rate (flows/s), the load variable \\
$s_f$, $d_f$ & source/destination access satellites of flow $f$ \\
$V_f$, $V_f^{\mathrm{rem}}(t)$ & original and remaining flow volume (Mbit) \\
$\pi_f(t)$, $H_f(t)$ & path of flow $f$ and its hop count \\
$x_f(t)$, $\tilde x_f(t)$ & sending rate; credited delivery rate (Mbps) \\
$x_f^{\mathrm D}(t)$, $\alpha_f$ & DCTCP rate (Mbps); congestion estimate \\
$A_l(t)$, $\rho_l(t)$ & link arrival rate (Mbps); drop fraction of arrivals \\
$\xi_l(t)$ & link overflow volume per step (Mbit) \\
$m_l(t)$, $F_f(t)$ & link ECN marking probability; flow marked fraction \\
$R_l(t)$ & explicit rate advertised by link $l$ (Mbps) \\
$d_{\mathrm g}$ & one-way delay per ground access link (s) \\
$d_f^{\mathrm{prop}}(t)$ & ISL propagation delay along the path of $f$ (s) \\
$\mathrm{RTT}_f(t)$ & effective RTT of flow $f$ in the end-host model (s) \\
$D_f(t)$ & one-way delay of flow $f$ including access links (s) \\
$Q_f(t)$ & queueing delay along the path of flow $f$ (s) \\
$\Theta$ & aggregate credited throughput over an episode (Mbps) \\
$D_q$ & rate-weighted mean queueing delay (s) \\
$\rho_{\mathrm{drop}}$ & episode drop rate, including link-down losses \\
$\mathrm{hop}(i,d)$ & shortest hop count from satellite $i$ to destination $d$ \\
$h^{+}_{i,d,p}$ & extra hops over a shortest path via port $p$ toward $d$ \\
$\Adm_{i,d}(k)$ & admissible ports of satellite $i$ toward $d$ \\
$W_{i,d,p}(k)$ & normalized slow-layer base weight of port $p$ toward $d$ \\
$\varphi_{i,d,p}(t)$ & unnormalized fast-layer forwarding weight of port $p$ \\
$o_i(k)$, $r(k)$ & local observation of satellite $i$; reward in period $k$ \\
$\pi_i$, $\mathcal J(\pi;\vartheta)$ & slow-layer policy of $i$; expected mean episode reward \\
$\vartheta$ & fast-layer configuration (mechanisms and parameters) \\
$\kappa$ & queue sensitivity of fast-layer load balancing \\
$\eta$ & RCP target utilization \\
$u_{i,d,p}(k)$ & previous-period downstream occupancy via $p$ toward $d$ \\
$a_1$, $a_2$ & occupancy thresholds for fairly busy and busy ports \\
$\chi_1$, $\chi_2$ & fractions diverted from fairly busy and busy ports \\
$\epsilon$ & weight floor of the soft split \\
\bottomrule
\end{tabularx}
\end{table}

\subsection{Constellation and Topology}
\label{sec:topology}
The constellation is a Walker-Delta $53^\circ{:}\,144/12/1$ configuration at an altitude of 550\,km, i.e., $N_{\mathrm o}=12$ orbital planes with $N_{\mathrm s}=12$ satellites each and $N=N_{\mathrm o}N_{\mathrm s}=144$.
Satellites move on circular orbits under two-body dynamics around a spherical Earth of radius 6371\,km that rotates at its sidereal rate.
Satellite $j\in\{0,\dots,N_{\mathrm s}-1\}$ of plane $o\in\{0,\dots,N_{\mathrm o}-1\}$ has the right ascension of the ascending node $2\pi o/N_{\mathrm o}$ and the initial argument of latitude $2\pi j/N_{\mathrm s}+2\pi o/N$ (phasing factor 1).
We denote by $\mathbf x_i(t)$ and $\phi_i(t)$ the Earth-fixed position and the latitude of satellite $i$.

Every satellite has four ISL ports $p\in\mathcal P=\{1,2,3,4\}$ in the +Grid pattern~\cite{bhattacherjee2019network}, and $n(i,p)$ denotes the neighbor behind port $p$ of satellite $i$.
Ports 1 and 2 connect to the following and the preceding satellite in the same plane.
Ports 3 and 4 connect to satellites in the eastern and the western adjacent plane: each satellite is paired with the satellite of the adjacent plane whose argument of latitude is closest at $t=0$, and the pairing is kept fixed afterward.
Every pair of neighbors is connected by two directed ISLs $l=(i,p)$, which form the link set $\mathcal L$ with $|\mathcal L|=4N$.
Because the relative angular velocity between adjacent planes is large near the latitude extremes of an inclined orbit, an inter-plane ISL is switched off whenever either endpoint is at or above $\phi_{\max}=45^\circ$ in absolute latitude.
This is the main source of periodic topology change.
Link $l=(i,p)$ is available if $e_l(t)=e^{\mathrm{orb}}_l(t)\bigl(1-\zeta_l(t)\bigr)=1$, where the orbital availability is
\begingroup\predisplaypenalty=0
\begin{equation}
e^{\mathrm{orb}}_l(t)=
\begin{cases}
1, & p\in\{1,2\},\\
\mathbf 1\bigl[\max_{j\in\{i,\,n(i,p)\}}|\phi_j(t)|<\phi_{\max}\bigr], & p\in\{3,4\},
\end{cases}
\label{eq:avail}
\end{equation}
\endgroup
and $\zeta_l(t)\in\{0,1\}$ indicates a link outage due to a random failure in the stress setting described below or an injected failure in the transient experiment.
The available links form the time-varying directed graph $\mathcal G(t)=(\Sat,\{l\in\mathcal L: e_l(t)=1\})$.
Each directed ISL has capacity $C_l=100$\,Mbps and propagation delay $d_l(t)=\|\mathbf x_i(t)-\mathbf x_{n(i,p)}(t)\|/c_0$, where $c_0$ is the speed of light.
Positions, link states, and delays are refreshed every $\tau_{\mathrm{top}}$ and held constant in between.

We denote by $\mathrm{hop}(i,d)$ the shortest hop count from satellite $i$ to satellite $d$ on the current graph $\mathcal G(t)$; it corresponds to the link-state information that a routing protocol already distributes.
We assume that link-state information reaches every satellite within $\tau_{\mathrm{top}}$.
We suppress the time argument of $\mathrm{hop}$, set $\mathrm{hop}(i,i)=0$, and treat an unreachable destination separately, without evaluating differences of infinite hop counts.
For a reachable destination $d\ne i$ and an available port $p$, forwarding through $p$ and then following a shortest path reaches $d$ with
\begin{equation}
h^{+}_{i,d,p}=1+\mathrm{hop}\bigl(n(i,p),d\bigr)-\mathrm{hop}(i,d)
\label{eq:extra}
\end{equation}
extra hops.
We call the available ports with $h^{+}_{i,d,p}=0$, which lie on a shortest path, the \emph{shortest-direction} ports of $i$ toward $d$.
Because the hop counts of neighboring satellites differ by at most one, $h^{+}_{i,d,p}\in\{0,1,2\}$.
The quantity $h^{+}_{i,d,p}$ describes a local next-hop choice; extra hops can accumulate along a path.

Under real orbital dynamics, a default episode of 30\,s covers only about 0.5\% of an orbit.
To also test faster topology change, we use a \emph{stress} setting in which orbital motion and Earth rotation are accelerated $20\times$ and ISLs fail at random.
At every topology refresh, each available undirected ISL fails with probability $\nu\tau_{\mathrm{top}}$, where $\nu=0.002$\,s$^{-1}$, and then stays down in both directions for an exponentially distributed repair time of mean 5\,s.
Unlike orbit-driven switch-offs, these random failures cannot be predicted.

\subsection{Ground Access and Traffic}
\label{sec:traffic}
Flow endpoints are drawn from a set $\mathcal C$ of 30 metropolitan hotspots with weights $\omega_c$ approximately proportional to their population, $\sum_{c\in\mathcal C}\omega_c=1$.
The source hotspot of a flow is $c$ with probability $\omega_c$, and its destination hotspot is $c'\neq c$ with probability $\omega_{c'}/(1-\omega_c)$.
Each ground endpoint is placed around its hotspot with a Gaussian jitter of $1.5^\circ$ in latitude and longitude.
Flows arrive as a Poisson process with rate $\lambda$~flows/s, which we use as the load variable.
The size of flow $f$ is $V_f=\min\{V,V_{\max}\}$, where $V$ follows a Pareto distribution~\cite{crovella1997self} with $\Pr[V>v]=(v/V_{\min})^{-1.2}$ for $v\ge V_{\min}=0.5$\,Mbit, and $V_{\max}=200$\,Mbit.
We call flows smaller than 1\,Mbit \emph{mice} and use their FCT to measure short-flow performance.

Each ground endpoint accesses the satellite nearest to it in Euclidean distance; we impose no minimum elevation angle.
We denote by $s_f$ and $d_f$ the current source and destination satellites of flow $f$, suppressing their time arguments.
They are chosen when the flow arrives and re-evaluated at the end of every control period; when one of them changes, the flow is reassociated and re-routed.
The ground--satellite segment is modeled by a fixed one-way delay of $d_{\mathrm g}=4$\,ms per access link and a per-flow access rate limit of 20\,Mbps.
Access links have neither a queue nor a shared capacity in our model, so congestion arises only on ISLs.

\subsection{Flow-Level Data Plane}
\label{sec:dataplane}
\textbf{Paths.}
Flow $f$ follows a loop-free path $\pi_f(t)=(l_{f,1},\dots,l_{f,H_f(t)})$ of directed ISLs from $s_f$ to $d_f$ with at most $H_{\max}$ hops ($32$ by default).
Paths are chosen per flowlet.
Because a fluid model has no packet gaps, each active flow independently starts a new flowlet in each step with probability $\Delta t/\tau_{\mathrm{fl}}$, where $\tau_{\mathrm{fl}}=0.1$\,s.
This gives geometrically distributed intervals with mean $\tau_{\mathrm{fl}}$ and approximates Poisson flowlet boundaries in discrete time.
At a flowlet boundary, and when a new flow arrives, the path is drawn hop by hop.
At satellite $i$, the next port is sampled with probability proportional to the forwarding weight $\varphi_{i,d_f,p}(t)$ among the available ports that lead to unvisited satellites.
If all these ports have zero weight, the walk takes the one with the smallest hop count to $d_f$; if none of them exists (a dead end) or the walk exceeds $H_{\max}$ hops, the flow falls back to a shortest-hop path.
When a link on the path is switched off or fails, or an endpoint is reassociated, the flow is re-routed immediately; we count these \emph{forced re-routes} separately.
A flow with no valid path, including one whose shortest-path fallback cannot reach the destination within $H_{\max}$ hops, sends nothing and retries in every step.
A same-satellite flow ($s_f=d_f$) has a valid empty path and generates no ISL arrivals.
For an empty path, sums are zero, products are one, and the path rate limit $\min_{l\in\pi_f(t)}R_l(t)$ is $\infty$.
Except for forced re-routes, existing flows with a valid path keep their paths until their next flowlet boundary, so a new split of the slow layer reaches each flow at its next flowlet.

\textbf{Fluid queues.}
In step $t$, flow $f$ sends at rate $x_f(t)$, set by its end host.
Its fluid reaches hop $h$ reduced by the losses at the preceding hops,
\begin{equation}
x_{f,h}(t)=x_f(t)\prod_{h'=1}^{h-1}\bigl(1-\rho_{l_{f,h'}}(t-1)\bigr),
\label{eq:hoprate}
\end{equation}
where $\rho_l(t)$ is the drop fraction of link $l$ defined below; the one-step lag avoids solving a fixed point within a step.
The arrival rate at link $l$ is $A_l(t)=\sum_{f}\sum_{h:\,l_{f,h}=l}x_{f,h}(t)$.
The model applies arrivals at all hops in the same step, so downstream queues can respond to rate and path changes before the traffic would physically arrive.
Propagation delays enter only the RTT and completion-time calculations.
We model the first-in, first-out (FIFO) buffer of each directed ISL as an aggregate fluid queue of size $B=5$\,Mbit.
At satellite $i$ in Fig.~\ref{fig:system}, the output queue at the eastward port $p=3$ is 80\% full, compared with 20\% at the in-plane port $p=1$.
Here $q_l(t)$ is the queue immediately before the fluid update in step $t$, and $q_l(t+1)$ is its value after that update:
\begin{align}
q_l(t+1)&=\min\Bigl\{\bigl[q_l(t)+\bigl(A_l(t)-e_l(t)C_l\bigr)\Delta t\bigr]^{+},\,B\Bigr\},\label{eq:queue}\\
\xi_l(t)&=\bigl[q_l(t)+\bigl(A_l(t)-e_l(t)C_l\bigr)\Delta t-B\bigr]^{+},\label{eq:overflow}
\end{align}
where the overflow $\xi_l(t)$ is dropped, and the drop fraction is $\rho_l(t)=\xi_l(t)/(A_l(t)\Delta t)$ if $A_l(t)>0$ and $\rho_l(t)=0$ otherwise.
Initially, $q_l(0)=0$ and $\rho_l(-1)=0$.
In each step, the model assigns flow $f$ a credited delivery rate of
\begin{equation}
\tilde x_f(t)=\Bigl[x_f(t)-\sum_{h=1}^{H_f(t)}x_{f,h}(t)\,\rho_{l_{f,h}}(t)\Bigr]^{+},
\label{eq:delivered}
\end{equation}
and its remaining volume $V^{\mathrm{rem}}_f(t)$, initialized to $V_f$, decreases by $\tilde x_f(t)\Delta t$ in every step, so that lost data stay in the remaining volume and are retransmitted.
With the one-step lag in Eq.~\eqref{eq:hoprate} and these current-step loss deductions, end-to-end volume is conserved only approximately within each step during transients.
When a link goes down, the model clears its queue and counts the discarded volume as lost.
Before re-routing, the model adds each active flow's share of this loss to its remaining volume, capped at $V_f$, if the flow's path used the link when it went down.
This share is the flow's fraction of arrivals to the link in the preceding step, or zero if the link received no traffic in that step.
Only queue overflow and link outages cause losses in the model.

\textbf{End hosts.}
ECN marking and the effective RTT for the DCTCP update use the queues after the current fluid update.
Each link marks ECN with probability $m_l(t)=\min\{1,[(q_l(t+1)-K_{\mathrm{ecn}})/\delta]^{+}\}$ with a fixed threshold $K_{\mathrm{ecn}}=0.5$\,Mbit and smoothing width $\delta=0.05$\,Mbit, so that the fraction $F_f(t)=1-\prod_{l\in\pi_f(t)}\bigl(1-m_l(t)\bigr)$ of the traffic of flow $f$ is marked.
End hosts use a DCTCP-inspired fluid rate model with stepwise ECN feedback~\cite{alizadeh2010dctcp}.
Acknowledgments are not modeled; the effective RTT used by the end-host model for flow $f$ is
\begin{equation}
\mathrm{RTT}_f(t)=2\bigl(d^{\mathrm{prop}}_f(t)+d_{\mathrm g}\bigr)+\sum_{l\in\pi_f(t)}q_l(t+1)/C_l,
\label{eq:rtt}
\end{equation}
where $d^{\mathrm{prop}}_f(t)=\sum_{l\in\pi_f(t)}d_l(t)$ is the ISL propagation delay of its path.
The simulations use $2d_{\mathrm g}$ as an approximate ground-access contribution to this control timescale; a physical round trip through both access links would contribute $4d_{\mathrm g}$.
Unless stated otherwise, the fluid implementation applies the current ECN fraction and advertised rate limits without a separate feedback propagation delay; the RTT scales the end-host updates below.
Physical feedback reaches the sender about one RTT later.
In a delayed variant, the feedback lag in data steps is $\lceil\mathrm{RTT}_f/\Delta t\rceil$, computed using the pre-update queues $q_l(t)$ and clipped to 1--127 steps (5--635\,ms), and the sender uses the marked fraction $F_f$, the path rate limit $\min_{l\in\pi_f}R_l$, and the loss signal from that many steps earlier. The ECN and loss histories start at zero, and the rate-cap history is initialized from the first valid path's advertised cap.
After the fluid update in step $t$, the model updates the congestion estimate as $\alpha_f\leftarrow\alpha_f+g\,(\Delta t/\mathrm{RTT}_f)(F_f-\alpha_f)$ with $g=1/16$ and computes the DCTCP rate for step $t+1$ as
\begin{equation}
x^{\mathrm D}_f \leftarrow x^{\mathrm D}_f+\frac{\Delta t}{\mathrm{RTT}_f}\Bigl(\frac{M}{\mathrm{RTT}_f}-\frac{x^{\mathrm D}_f\,\alpha_f F_f}{2}\Bigr),
\label{eq:dctcp}
\end{equation}
with segment size $M=12$\,kbit. An overflow loss or a link going down on the flow's path triggers rate halving, at most once per RTT. In the default variant, link-down signals are applied before re-routing and sending, and overflow signals are applied after the fluid update. The delayed variant applies historical loss signals after the fluid update.
The congestion estimate $\alpha_f$ starts at zero, and the DCTCP rate $x^{\mathrm D}_f$ starts at 10\,Mbps and is kept within $[0.1,20]$\,Mbps, whose upper end is the access rate limit.
The sending rate is
\begin{equation}
x_f(t)=\min\Bigl\{x^{\mathrm D}_f(t),\ \frac{V^{\mathrm{rem}}_f(t)}{\Delta t},\ \min_{l\in\pi_f(t)}R_l(t)\Bigr\},
\label{eq:rate}
\end{equation}
where $R_l(t)$ is the explicit rate advertised by link $l$ if the fast layer applies explicit-rate control and $R_l\equiv\infty$ otherwise.
Equation~\eqref{eq:rate} applies to flows with a valid path; a flow with no valid path has $x_f(t)=0$.

\textbf{Delays and completion times.}
The one-way delay of flow $f$, including the uplink and the downlink, is
\begin{equation}
D_f(t)=d^{\mathrm{prop}}_f(t)+2d_{\mathrm g}+\sum_{l\in\pi_f(t)}q_l(t+1)/C_l .
\label{eq:owd}
\end{equation}
Under this flow-level accounting, the FCT of a flow is estimated as the time from its arrival until its remaining volume reaches zero, plus its one-way delay at that moment.

\subsection{Performance Metrics}
\label{sec:metrics}
Consider an episode of $K$ control periods, let $\mathcal T$ be the set of its steps, and let $Q_f(t)=\sum_{l\in\pi_f(t)}q_l(t+1)/C_l$ be the queueing delay along the path of flow $f$, sampled after the fluid update.
We report the throughput $\Theta$, the queueing delay $D_q$ weighted by the sending rate, and the drop rate $\rho_{\mathrm{drop}}$,
\begin{align}
\Theta&=\frac{1}{KT_c}\sum_{t\in\mathcal T}\sum_{f}\tilde x_f(t)\,\Delta t,\label{eq:thr}\\
D_q&=\frac{\sum_{t\in\mathcal T}\sum_{f}x_f(t)\,Q_f(t)}{\sum_{t\in\mathcal T}\sum_{f}x_f(t)},\label{eq:qdelay}\\
\rho_{\mathrm{drop}}&=\frac{\sum_{t\in\mathcal T}\sum_{l\in\mathcal L}\xi_l(t)+V^{\mathrm{down}}}{\sum_{t\in\mathcal T}\sum_{f}x_f(t)\,\Delta t},\label{eq:drop}
\end{align}
where $V^{\mathrm{down}}$ is the volume lost when links go down.
Data credited before a link outage retain that credit and can count again upon retransmission, so $\Theta$ measures throughput under this accounting convention rather than receiver-side goodput.
The implementation accumulates the queueing-delay and drop ratios using data volumes, with a denominator floor of $10^{-9}$\,Mbit.
Link-down losses can include data queued in earlier steps; the episode drop metric is not clipped.
A flow counts as completed if its remaining volume reaches zero within the episode.
For completed mice, we report the mean and 99th-percentile FCT. If no mice complete, both values are recorded as not a number (NaN).
Elephants are flows of at least 20\,Mbit.
A flow is eligible if its destination is reachable at arrival and it can finish sending by the end of the episode under the ideal reference.
The reference uses the path with the smallest propagation delay at arrival and assumes no queueing, loss, or explicit rate cap.
The reference sender starts at 10\,Mbps and follows the unmarked update in Eq.~\eqref{eq:dctcp} up to the access limit of 20\,Mbps.
The completion rate is the fraction of eligible flows that complete.
The path stretch is the traffic-weighted ratio of actual to shortest hop counts, $\sum_{t,f}x_f(t)H_f(t)/\sum_{t,f}x_f(t)\,\mathrm{hop}(s_f,d_f)$, over flows with a valid path and distinct, reachable source and destination satellites; it is set to zero when its denominator is zero.
Jain's fairness index~\cite{jain1984quantitative} $(\sum_{l}u_l)^2/(|\mathcal L'|\sum_{l}u_l^2)$ is taken over the set $\mathcal L'$ of links that are available at some time in the episode, where the utilization $u_l$ is the data served by link $l$ divided by $C_l$ times its available time.
The index is undefined (NaN) if this set is empty or all its utilizations are zero.
Finally, we count the forced re-routes.
The same quantities over the steps of a single period $k$ are written with the argument $k$, e.g., $\Theta(k)$.

\section{Problem Formulation}
\label{sec:problem}

\subsection{Control Variables and Timescales}
Control acts on two timescales, with the fast-layer mechanisms and their parameters collected in a configuration $\vartheta$.
At the beginning of period $k$, the slow layer of each satellite $i$ chooses, for every reachable destination satellite $d\ne i$, split weights $W_{i,d,p}(k)$ over its ports, which stay fixed during $[kT_c,(k+1)T_c)$.
For slow-layer decisions, reachability, link availability, and hop counts refer to the latest link-state snapshot available before the period-boundary topology refresh, which is $\tau_{\mathrm{top}}$ old except at initialization.
Whenever a path is selected or reselected (Section~\ref{sec:dataplane}), the fast layer applies a fixed rule to the base split and local queues to obtain the forwarding weights,
\begin{equation}
\varphi_{i,d,p}(t)=\Phi_p\bigl(W_{i,d,\cdot}(k),\,q_{i,\cdot}(t);\vartheta\bigr),
\label{eq:fwd}
\end{equation}
where $q_{i,\cdot}(t)$ denotes the instantaneous output queues at satellite $i$ and $\varphi_{i,d,p}(t)=0$ whenever $W_{i,d,p}(k)=0$.
When explicit-rate control is enabled, the fast layer updates $R_l(t)$ in Eq.~\eqref{eq:rate} every $\tau_f$ using the current queue at link $l$ and its arrival rate over the preceding fast interval.
Without a fast layer, $\varphi_{i,d,p}(t)=W_{i,d,p}(k)$ and $R_l\equiv\infty$.
For any such fast layer, P1 optimizes the slow-layer policy with $\vartheta$ fixed.

\textbf{Information structure.}
To keep execution distributed, the decision of satellite $i$ for period $k$ may depend only on information available on board at the beginning of the period: the current state and the statistics over the previous period of its own ports; statistics of its neighbors from earlier periods, which neighbors exchange once per period; the topology and hop counts obtained from link-state information; orbit-predicted satellite positions and the time until each of its links is switched off by Eq.~\eqref{eq:avail}, within a horizon of $M_\ell=8$ periods (4\,s); and the traffic that it sources, sinks, and forwards.
We collect this information in the local observation $o_i(k)$; all statistics from earlier periods are zero at the beginning of an episode.
A policy may use any part of $o_i(k)$.

\subsection{Admissible Ports}
At the beginning of period $k$, satellite $i$ restricts the support of its base split toward each reachable destination $d\ne i$ to the admissible set
\begin{equation}
\begin{aligned}
\Adm_{i,d}(k)=\bigl\{p:\ & e_{(i,p)}=1,\ \text{and}\\
& h^{+}_{i,d,p}=0 \text{ if } \mathrm{hop}(i,d)=1\bigr\}.
\end{aligned}
\label{eq:adm}
\end{equation}
All quantities in Eq.~\eqref{eq:adm} use the pre-refresh snapshot, with $h^{+}_{i,d,p}$ computed by Eq.~\eqref{eq:extra}.
Only the direct port to $d$ is admissible when $\mathrm{hop}(i,d)=1$. Otherwise, every available port is admissible.
This constraint applies to all slow-layer splits.
Whole paths are subject to the loop-free and $H_{\max}$ conditions of Section~\ref{sec:dataplane}.
The set is locally computable and contains all shortest-direction ports, so it is non-empty for any reachable $d\ne i$.
No forwarding split is needed for $d=i$; for destinations unreachable in this snapshot, the stored weights are zero.
The simulator makes the updated graph available to the data plane at each topology refresh.
Weighted sampling masks links unavailable on that graph. A port unavailable in the pre-refresh snapshot keeps zero base weight until the next slow update.
The fallback rules of Section~\ref{sec:dataplane} use the current graph and may select ports outside the period-start admissible set, including for destinations that become reachable before the next slow update.

\subsection{Network Objective}
The reward in period $k$ combines a network-wide term with per-satellite terms averaged over all satellites,
\begin{equation}
r(k)=(1-\alpha)\, r^{\mathrm{glob}}(k)+\frac{\alpha}{N}\sum_{i\in\Sat} r^{\mathrm{loc}}_i(k),\qquad \alpha=0.3,
\label{eq:reward}
\end{equation}
with
\begin{align}
r^{\mathrm{glob}}(k) &= \frac{\Theta(k)}{\Theta_0}-\frac{D_q(k)}{D_0}-2\rho_{\mathrm{drop}}(k)-0.5\,\rho_{\mathrm{fr}}(k)\notag\\
&\quad -w_{\mathrm{fct}}\,\overline{\mathrm{FCT}}_{\mathrm{mice}}(k)/\mathrm{FCT}_0,\label{eq:rglob}\\
r^{\mathrm{loc}}_i(k) &= -\bar{u}^{q}_i(k)-2\,\rho^{\mathrm{drop}}_i(k) .\label{eq:rloc}
\end{align}
Here $\Theta(k)$, $D_q(k)$, and $\rho_{\mathrm{drop}}(k)$ are the metrics in Eqs.~\eqref{eq:thr}--\eqref{eq:drop} over period $k$, with $\rho_{\mathrm{drop}}(k)$ capped at one.
The reference scales are $\Theta_0=2000$\,Mbps and $D_0=0.05$\,s.
The forced re-route ratio $\rho_{\mathrm{fr}}(k)$ is the number of forced re-routes in period $k$ divided by the number of distinct flows active at its start or arriving during it.
Repeated forced re-routes of the same flow are counted separately, so the ratio can exceed one; it is zero if no flow is present.
The term $\overline{\mathrm{FCT}}_{\mathrm{mice}}(k)$ is the mean FCT of the mice completed in period $k$ (zero if none completes), with $\mathrm{FCT}_0=0.2$\,s and $w_{\mathrm{fct}}=1$.
The local term of satellite $i$ penalizes the mean buffer occupancy $\bar{u}^{q}_i(k)$, i.e., $q_{i,p}(t+1)/B$ averaged first over the steps of the period and then over the ports of satellite $i$ available at its end (zero if none is available).
The local drop rate $\rho^{\mathrm{drop}}_i(k)$ is the total volume lost at all outgoing ports of $i$, including queue overflow and link-down losses, divided by their total arrival volume in the period; it is capped at one and set to zero if the arrival volume is at most $10^{-9}$\,Mbit.
The reference scales make the reward dimensionless.

We seek a slow-layer policy $\pi=(\pi_i)_{i\in\Sat}$ that maximizes the expected mean reward over an episode of $K$ periods ($K=60$ by default).
\begin{subequations}
\label{eq:P1}
\begin{align}
\textbf{P1:}\ \max_{\pi}\ \ & \mathcal J(\pi;\vartheta)=\mathbb E\Bigl[\frac{1}{K}\sum_{k=0}^{K-1}r(k)\Bigr] \label{eq:P1-obj}\\
\text{s.t.}\ \ & W_{i,d,p}(k)\ge 0,\quad \textstyle\sum_{p}W_{i,d,p}(k)=1, \label{eq:P1-simplex}\\
& W_{i,d,p}(k)=0,\quad \forall p\notin\Adm_{i,d}(k), \label{eq:P1-adm}\\
& W_{i,\cdot,\cdot}(k)=\pi_i\bigl(o_i(k)\bigr), \label{eq:P1-info}
\end{align}
\end{subequations}
for all satellites $i$, destinations $d\ne i$ reachable in the snapshot used for period $k$, ports $p$, and periods $k$, subject to the data plane in Eqs.~\eqref{eq:hoprate}--\eqref{eq:owd} driven by the fast layer in Eq.~\eqref{eq:fwd}.
The expectation is over flow arrivals, endpoints, and sizes; flowlet boundaries; path sampling; and, in the stress setting, link failures.
We call $\mathcal J$ the \emph{evaluation reward} and use it, together with the metrics of Section~\ref{sec:metrics}, to compare all methods.

P1 is difficult to solve exactly, for three reasons.
First, the split $W(k)$ affects the reward only through the closed loop of path sampling, end-host reaction, and the fast layer, and through flows that arrive later; the dynamics in Eqs.~\eqref{eq:hoprate}--\eqref{eq:rate} are nonlinear and stochastic, and the FCT term couples all hops of a flow.
Second, the information constraint in Eq.~\eqref{eq:P1-info} makes P1 a decentralized problem under partial observability: every satellite sees a local and partly delayed view, while the splits of all satellites interact on shared links.
Third, the topology, the access satellites, and, in the stress setting, the link states change within an episode, so the slow layer must adapt to time-varying forwarding conditions.

We therefore compare parameterized distributed slow-layer policies that satisfy Eqs.~\eqref{eq:P1-simplex}--\eqref{eq:P1-info} by construction.
We select free parameters on validation seeds and evaluate $\mathcal J$ on separate test seeds.
The main comparison applies the same fast-layer rules to all methods, with a separate $\kappa$ selected to maximize validation $\mathcal J$ for each method that uses queue reweighting.
The centralized MCF reference uses global demand and lies outside P1.
The slow-layer split determines which ports the fast layer can sample.

\section{Proposed Method}
\label{sec:method}

\subsection{Overview}
\begin{figure}[!t]
\centering
\includegraphics[width=\columnwidth]{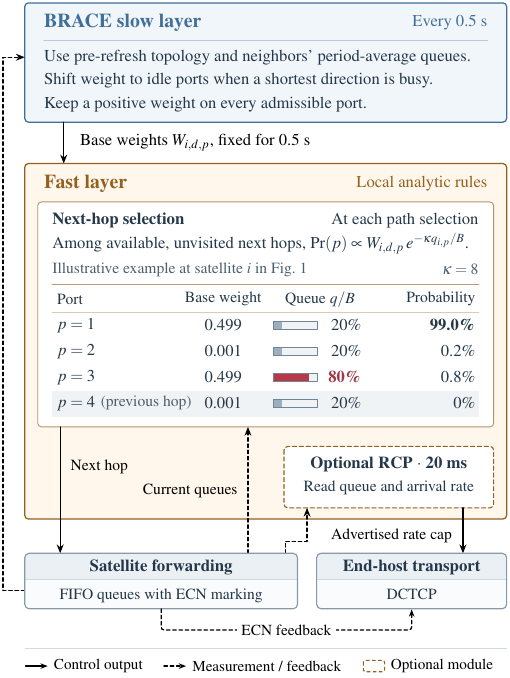}
\caption{Dual-timescale control in BRACE.}
\label{fig:arch}
\end{figure}

Fig.~\ref{fig:arch} shows the dual-timescale control architecture of BRACE.
At each control-period boundary, the slow layer uses the pre-refresh link-state snapshot to compute admissible ports via Eq.~\eqref{eq:adm} and assign destination-specific base weights.
At each path-selection event, the fast layer reweights the split using instantaneous local queues.
Sustained arrivals at twice an ISL's capacity can fill an empty buffer in $B/C_l=50$\,ms, a tenth of $T_c$.
An optional explicit-rate controller updates each port's advertised rate every $\tau_f$, with the minimum rate along a flow's path capping its sending rate.
Both layers use analytic rules.

The fast layer and the slow-layer interface form a framework in which the slow layer is a replaceable policy.
The slow layer of BRACE adapts the explicit load-balancing rule of ELB~\cite{taleb2009elb} and applies a weight floor to the resulting split; the ELB baseline uses the same rule without the floor.

\subsection{Fast Layer}
\textbf{Local load balancing.}
At each path-selection event, satellite $i$ samples a port $p$ toward $d$ among currently available ports leading to unvisited satellites with probability proportional to
\begin{equation}
W_{i,d,p}(k)\,\exp\bigl(-\kappa\, q_{i,p}(t)/B\bigr),
\label{eq:lb}
\end{equation}
where $\kappa$ is the queue sensitivity and $q_{i,p}(t)$ is the instantaneous queue of the port; this weight is the forwarding weight $\varphi_{i,d,p}(t)$ of Eq.~\eqref{eq:fwd}.
Because $W_{i,d,p}(k)=0$ outside $\Adm_{i,d}(k)$, weighted path sampling stays within the admissible set.
For $\kappa=\infty$, the rule chooses a minimum-queue port among currently available, unvisited ports with positive $W$, breaking queue ties in proportion to $W$.
In Fig.~\ref{fig:arch}, all four ports at satellite $i$ are admissible for the flow in Fig.~\ref{fig:system}.
With $\epsilon=10^{-3}$, the two shortest-direction ports have equal base weights because their downstream ports toward $d_f$ are nearly empty. The fast layer favors $p=1$ because its local queue is shorter than that of $p=3$, and masks $p=4$ because it leads to the previous hop.
Path selection uses the fallback rules of Section~\ref{sec:dataplane} when needed.
The rule follows the flowlet-based load balancers of data-center networks~\cite{alizadeh2014conga,vanini2017letflow,katta2016hula}, but uses only the local queue and takes its base split from the slow layer.

\textbf{Explicit-rate control.}
When explicit-rate control is enabled, each port initializes its advertised rate $R_l$ to $C_l$ and updates it every $\tau_f$ using an XCP/RCP-style rule~\cite{katabi2002xcp,dukkipati2006rcp}, referred to as RCP below,
\begin{align}
g_l &= \frac{\tau_f}{\bar d}\cdot\frac{a\,(\eta C_l-y_l)-b\,q_l/\bar d}{\eta C_l},\label{eq:rcp-gain}\\
R_l &\leftarrow \min\Bigl\{C_l,\ \max\bigl\{R_{\min},\ R_l\,[1+\mathrm{clip}(g_l,-0.9,1)]\bigr\}\Bigr\},\label{eq:rcp-rate}
\end{align}
where $y_l$ is the arrival rate at $l$ over the preceding fast interval, $\bar d=0.1$\,s is a nominal RTT, $a=0.4$, $b=0.2$, the target utilization is $\eta=1.0$, and $R_{\min}=0.5$\,Mbps.
At each fast update, an unavailable port resets $R_l$ to $C_l$.
By Eq.~\eqref{eq:rate}, the advertised rate can cap a flow from its first sending step.
This optional controller complements traffic redistribution when the admissible ports cannot absorb the offered load.
In the high-load experiments, it reduces queueing delay and losses and increases the tail FCT of mice.
The main comparison enables it with the same parameters for all slow layers.

\subsection{Slow-Layer Interface}
\label{sec:interface}
The interface between the layers is the split table.
At the beginning of every period, a slow layer installs, for every reachable destination $d\ne i$, weights $W_{i,d,\cdot}(k)$ that satisfy Eqs.~\eqref{eq:P1-simplex} and~\eqref{eq:P1-adm} and stay fixed for $T_c$.
A distributed slow layer computes the table from $o_i(k)$ as required by Eq.~\eqref{eq:P1-info}.

As noted in Section~\ref{sec:problem}, a zero base weight excludes a port from weighted sampling.
BRACE therefore applies a \emph{weight floor} $\epsilon>0$ to the raw weights $\hat W_{i,d,p}$ of a slow layer and installs the \emph{soft} split
\begin{equation}
W_{i,d,p}=\frac{\mathbf 1[p\in\Adm_{i,d}]\bigl(\hat W_{i,d,p}+\epsilon\bigr)}{\sum_{p'\in\Adm_{i,d}}\bigl(\hat W_{i,d,p'}+\epsilon\bigr)},
\label{eq:analytic}
\end{equation}
which gives every admissible port a positive base weight. An admissible port can participate in weighted sampling while its link is available and its next satellite is unvisited.
The floor does not depend on how $\hat W$ is computed; the centralized reference also applies it.
For a raw split normalized over the admissible ports, a zero raw weight becomes $\epsilon/(1+|\Adm_{i,d}|\epsilon)\approx\epsilon$ for small $\epsilon$.
For $0<\kappa<\infty$, consider two currently available ports $p$ and $p'$ leading to unvisited satellites and having positive base weights.
By Eq.~\eqref{eq:lb}, $p'$ is at least as likely as $p$ when
\begin{equation}
q_{i,p}(t)-q_{i,p'}(t)\ \ge\ \frac{B}{\kappa}\,\ln\frac{W_{i,d,p}}{W_{i,d,p'}}.
\label{eq:escape}
\end{equation}
For $W_{i,d,p}=0.5$ and zero raw weight at $p'$, the queue difference is about $0.78B$ for $\kappa=8$ and $\epsilon=10^{-3}$, and about $0.14B$ for $\kappa=128$ and $\epsilon=10^{-8}$.
At $\kappa=\infty$, a shorter queue ranks $p'$ ahead of $p$ among eligible ports.
For fixed raw weights, increasing $\kappa$ or $\epsilon$ reduces the queue gap needed for a port with zero raw weight to match a preferred port's selection probability. We select the two parameters jointly on validation seeds.
In the mechanism studies with a common fast-layer setting, the floor reduces losses at high load with little change in mean reward.

\subsection{Analytic Slow Layer}
\label{sec:analytic}
Unlike the original ELB, in which satellites report congestion on state transitions, BRACE exchanges previous-period port occupancies with neighbors once per period.
Of the local observation $o_i(k)$ in Section~\ref{sec:problem}, BRACE uses only port occupancies and link-state information.
Let $j$ be the neighbor behind port $p$ of satellite $i$, and let
\begin{equation}
u_{i,d,p}(k)=\min_{p':\,h^{+}_{j,d,p'}=0}\ \bar q_{j,p'}(k-1)/B
\label{eq:downstream}
\end{equation}
be the minimum previous-period mean occupancy among $j$'s shortest-direction ports toward $d$ ($u_{i,d,p}=0$ if $j=d$).
Here $\bar q_{j,p'}(k-1)$ averages the post-update queues over period $k-1$, and shortest directions use the slow layer's link-state snapshot.
Each port is \emph{idle} if $u_{i,d,p}<a_1$, \emph{fairly busy} if $a_1\le u_{i,d,p}<a_2$, and \emph{busy} otherwise.
Starting from the even split over the shortest-direction ports $\mathcal S_{i,d}=\{p\in\Adm_{i,d}:h^{+}_{i,d,p}=0\}$, satellite $i$ removes the fraction $\chi_1$ ($\chi_2$) of the share of every fairly busy (busy) port in $\mathcal S_{i,d}$ and redistributes the removed traffic over the idle ports of $\Adm_{i,d}$ in proportion to $e^{-h^{+}_{i,d,p}}$, which favors shortest directions over detours; if no admissible port is idle, nothing is moved.
Denoting the result by $W^{\mathrm{an}}_{i,d,p}$, the slow layer outputs the soft split in Eq.~\eqref{eq:analytic} with $\hat W_{i,d,p}=W^{\mathrm{an}}_{i,d,p}$.
The thresholds and fractions, $a_1=5\times10^{-4}$, $a_2=0.01$, $\chi_1=0.2$, and $\chi_2=0.3$, are chosen by a grid search on validation seeds disjoint from the test seeds. The threshold $a_1$, the smallest value in the grid, corresponds to 2.5\,kbit, less than one segment, so even a very small average downstream queue moves a port out of the idle class.
Algorithm~\ref{alg:brace} summarizes the event order. The slow-layer and port operations run independently on each satellite.

\begin{algorithm}[!t]
\DontPrintSemicolon
\SetAlgoVlined
\caption{BRACE with an analytic slow layer}
\label{alg:brace}
\KwIn{Link-state snapshots of $\mathcal G(t)$, capacities $C_l$, queues $q_l(t)$, neighbor queue averages $\bar q_{j,p'}(k-1)$, and flow events with endpoints $(s_f,d_f)$.}
\KwOut{Base port weights $W_{i,d,p}(k)$, selected flow paths $\pi_f(t)$, and advertised link rates $R_l(t)$.}
Zero all statistics. Set $R_l=C_l$ if RCP is enabled and $R_l=\infty$ otherwise\;
For path selection, use weights $\varphi_{i,d_f,p}(t)$ from Eq.~\eqref{eq:lb} with the masks and fallback in Section~\ref{sec:dataplane}\;
\For{each data step $t=0,1,\dots$}{
  \If{control period $k$ begins}{
    Exchange port averages and clear the split table\;
    \ForEach{$d\ne i$ reachable in the pre-refresh snapshot}{
      Compute $\Adm_{i,d}(k)$ and $u_{i,d,p}(k)$ via Eqs.~\eqref{eq:adm} and~\eqref{eq:downstream}\;
      Classify ports using $a_1,a_2$ and split weight equally over the shortest directions $\mathcal S_{i,d}$\;
      \If{idle admissible ports exist}{
        Divert $\chi_1$ (fairly busy) or $\chi_2$ (busy) of each affected share to idle admissible ports in proportion to $e^{-h^{+}_{i,d,p}}$\;
      }
      Apply the floor $\epsilon$ and install $W_{i,d,p}(k)$ via Eq.~\eqref{eq:analytic}\;
    }
    Re-route reassociated flows\;
  }
  \If{an enabled rate update is due}{
    Update and advertise $R_l(t)$ on available links using $q_l(t)$ and $y_l$ in Eqs.~\eqref{eq:rcp-gain}--\eqref{eq:rcp-rate}\;
    Set $R_l(t)\gets C_l$ on unavailable links and clear fast-interval statistics\;
  }
  \If{a topology refresh is due}{
    Refresh topology. Clear downed queues, record losses, add loss shares to remaining volumes, and re-route the affected flows (Section~\ref{sec:dataplane})\;
  }
  Select $\pi_f(t)$ for arrivals, then new flowlets, then retries\;
  Compute $x_f(t)$ via Eq.~\eqref{eq:rate}, advance the data plane, and accumulate statistics\;
  \If{a slow control period ends}{
    Reassociate endpoints and flag affected flows for re-routing\;
    Save and reset period statistics\;
  }
}
\end{algorithm}

\subsection{Algorithmic Complexity}
\label{sec:complexity}
We count BRACE's control operations in Algorithm~\ref{alg:brace} given the hop counts maintained by the link-state routing protocol (Section~\ref{sec:topology}).
Let $P=|\mathcal P|=4$ be the number of ports per satellite.
For each destination, a satellite evaluates at most $P$ candidate ports.
Computing the downstream minimum in Eq.~\eqref{eq:downstream} directly requires at most $P$ occupancy comparisons per candidate.
Port classification, redistribution, and normalization in Eq.~\eqref{eq:analytic} then take $O(P)$ time for that destination.
With at most $N-1$ nonlocal destinations, a slow update thus takes $O(NP^2)$ time per satellite, and the split table occupies $O(NP)$ words.
With four ports, both costs are $O(N)$ per satellite and $O(N^2)$ across the constellation in the worst case.

The slow layer exchanges one average occupancy per local port with each neighbor every $T_c$.
This requires $P$ scalars per directed ISL per period, independent of the number of destinations.
Assuming constant-time visited-node tests, choosing a next hop by Eq.~\eqref{eq:lb} takes $O(P)$ time. A path selection, including at most one shortest-path fallback, requires $O(H_{\max}P)$ port checks.
The split-table bound counts only the base weights. The forwarding model assumes packets carry visited satellite identifiers and each satellite keeps one next-hop entry per active flowlet.
Given accumulated local measurements, the explicit-rate update in Eqs.~\eqref{eq:rcp-gain}--\eqref{eq:rcp-rate} takes $O(1)$ time and state per port every $\tau_f$, independent of the number of active flows.

\section{Performance Evaluation}
\label{sec:eval}

\subsection{Simulation Settings}
\label{sec:settings}
\subsubsection{Scenarios and Evaluation Protocol}
We simulate the system of Section~\ref{sec:system} with the parameters given there and in Section~\ref{sec:method}.
The main comparison uses episodes of 30\,s (60 control periods), all starting at $t=0$ with the same orbital phases and Earth rotation angle.
We evaluate every policy in seven hotspot load conditions, four under real orbital dynamics ($\lambda\in\{500,750,1000,1250\}$~flows/s) and three under the stress setting ($\lambda\in\{1000,1250,1500\}$), on five test seeds (101--105) that fix the traffic and the link failures for all methods.
The learned baseline MA-DRL is trained with three seeds (1--3); we report the mean and the standard deviation ($\pm$) over training seeds, each training seed first averaged over the test seeds.
An \emph{orbital-geometry control} uses 20 test seeds (101--120), each starting at an epoch drawn uniformly from one sidereal day and shared across methods, with $\kappa=8$ for all slow layers and $\epsilon=10^{-3}$ for BRACE.
We compare methods paired by test seed and call a difference \emph{consistent} if it has the same sign on all five test seeds (all 20 in the orbital-geometry control).
For MA-DRL, paired comparisons first average the three trained networks within each test seed.
Unless stated otherwise, figures show means over the five test seeds, with error bars spanning their range, and shading on load axes marks the stress setting.
Drop-rate and reward-difference axes that include zero use a symmetric logarithmic scale, which is linear near zero.
Parameters are selected on separate validation seeds (900001--900003).
All slow layers use the same fast-layer rules for local load balancing (LB) and RCP ($\eta=1.0$, $\tau_f=20$\,ms).
We evaluate each configuration using the reward defined in Section~\ref{sec:problem}.
The main comparison uses validation-selected parameters. For each baseline that uses queue reweighting, $\kappa$ is selected from eight values, $\{8,16,32,64,128,256,512,\infty\}$, to maximize the mean reward on the validation seeds across real loads 1000 and 1250 and stress load 1500.
BRACE uses the same validation criterion to compare 43 joint $(\kappa,\epsilon)$ settings, with $\kappa\in\{8,32,64,128,256\}$ and $\epsilon$ candidates from 0 (ELB's rule) to $10^{-2}$.
It selects $\kappa=128$ and $\epsilon=10^{-8}$, the smallest positive value in the grid.
The mechanism studies use a common-fast-layer reference ($\kappa=8$, $\epsilon=10^{-3}$ for BRACE) and change only the stated component or parameter.
The scalability and traffic-pattern experiments, along with the main transient runs, freeze each method's validation-selected parameters without retraining.

\subsubsection{Compared Slow Layers}
We compare BRACE with seven baseline designs (eight configurations, including a tuned variant of the congestion-aware heuristic), all using period-based splits and the same fast-layer rules.
\begin{itemize}
\item \emph{Shortest path (SP).}
SP uses one shortest-direction port with a fixed tie-break, preventing fast-layer redistribution.
\item \emph{Equal-cost multipath (ECMP)}~\cite{rfc2992}.
ECMP splits traffic evenly over all shortest-direction ports without using congestion information.
\item \emph{Congestion-aware heuristic (CA).}
CA weights each admissible port by $e^{-8\bar q_{i,p}/B}\,e^{-h^{+}_{i,d,p}}$, where $\bar q_{i,p}$ is the previous-period average queue of its own port.
For the tuned variant (CA-tuned), the coefficient of the detour penalty in the exponent increases from 1 to 2 based on validation, giving $e^{-2h^{+}_{i,d,p}}$.
\item \emph{ELB}~\cite{taleb2009elb}.
ELB uses the analytic rule in Section~\ref{sec:analytic} with the same validated thresholds and fractions as BRACE, but without the weight floor.
\item \emph{TLR}~\cite{song2014tlr}.
TLR classifies the larger of normalized local and downstream period-average buffer occupancies using thresholds 0.05 and 0.2, assigning shortest-direction weights 1, 0.5, and 0.3 to green, yellow, and red states.
It enables non-red admissible detours only when all shortest directions are red.
\item \emph{DBPR}~\cite{deng2023dbpr}.
Our period-based DBPR adaptation weights each satellite's per-destination backlog, estimated from its period-average ISL queues and destination traffic shares, by the straight-line propagation delay to the destination.
It splits traffic over shortest-direction ports, a restriction selected on the validation seeds, in proportion to the positive differences between the local and neighboring weighted backlogs.
\item \emph{MA-DRL}~\cite{lozano2025continual}.
The MA-DRL adaptation trains a shared Q-network offline with Double DQN and the original state encoding, with the fast layer and admissible-port constraints active, on stress-setting episodes without random failures at the three stress loads.
The queueing weight of the reward, the execution rule (softmax with temperature 100), and the checkpoint are selected on the validation seeds; the other reward terms and weights follow the original method.
During evaluation, the frozen network produces periodic traffic splits from admissible-port Q-values; online and federated updates are omitted.
\end{itemize}
As a reference outside the distributed setting, \emph{Centralized MCF} follows the global allocation of HGL~\cite{liu2017hgl}.
At the start of every period, it aggregates the demand $\min\{x^{\mathrm D}_f,V^{\mathrm{rem}}_f/T_c\}$ of all active flows between access satellites, with no collection or solver delay, and solves a destination-based multicommodity-flow linear program over the admissible ports with the piecewise-linear link cost of~\cite{fortz2000internet}.
The resulting flows are normalized into port fractions with BRACE's weight floor ($\epsilon=10^{-3}$) and run under the same fast layer, with $\kappa$ selected on the validation seeds as for the baselines.

For the scalability and traffic-pattern experiments, we linearly interpolate CA's log drop rates at $\kappa=8$ over the validation load grid to find loads that match target drop rates, then round the loads to the nearest 50~flows/s.

\begin{table*}[!t]
\centering
\caption{Reward and drop rate of slow layers under hotspot traffic.}
\label{tab:main}
\footnotesize
\setlength{\tabcolsep}{4pt}
\begin{threeparttable}
\begin{tabular*}{\textwidth}{@{\extracolsep{\fill}}lrrrrrrrr@{}}
\toprule
Slow layer & $\kappa$ & R500 & R750 & R1000 & R1250 & S1000 & S1250 & S1500 \\
\midrule
\multicolumn{9}{l}{\emph{Reward}} \\
\midrule
SP & -- & $-$0.212 & $-$0.317 & $-$1.242 & $-$2.686 & $-$0.892 & $-$1.921 & $-$3.633 \\
ECMP & 512 & $-$0.169 & $-$0.121 & $-$0.314 & $-$0.797 & $-$0.152 & $-$0.401 & $-$0.884 \\
CA & 64 & $-$0.232 & $-$0.098 & $-$0.073 & $-$0.133 & $-$0.098 & $-$0.186 & $-$0.385 \\
CA-tuned & 32 & $-$0.158 & $-$0.025 & 0.029 & $-$0.014 & 0.008 & $-$0.049 & $-$0.223 \\
TLR & $\infty$ & $-$0.168 & $-$0.104 & $-$0.130 & $-$0.245 & $-$0.087 & $-$0.194 & $-$0.423 \\
DBPR & 32 & $-$0.159 & $-$0.095 & $-$0.213 & $-$0.650 & $-$0.134 & $-$0.361 & $-$0.819 \\
MA-DRL & $\infty$ & $-$0.315{\scriptsize$\pm$0.003} & $-$0.198{\scriptsize$\pm$0.003} & $-$0.185{\scriptsize$\pm$0.003} & $-$0.278{\scriptsize$\pm$0.004} & $-$0.190{\scriptsize$\pm$0.007} & $-$0.291{\scriptsize$\pm$0.013} & $-$0.532{\scriptsize$\pm$0.021} \\
ELB & 512 & $-$0.153 & $-$0.027 & 0.063 & 0.049 & 0.052 & 0.039 & $-$0.103 \\
\midrule
\textbf{BRACE} & 128 & \textbf{$-$0.148} & \textbf{$-$0.006} & \textbf{0.081} & \textbf{0.071} & \textbf{0.065} & \textbf{0.052} & \textbf{$-$0.068} \\
\midrule
\textit{Centralized MCF}$^\dagger$ & 64 & $-$0.136 & 0.035 & 0.181 & 0.240 & 0.161 & 0.209 & 0.186 \\
\midrule
\multicolumn{9}{l}{\emph{Drop rate (\%)}} \\
\midrule
SP & -- & 0.012 & 0.107 & 0.372 & 2.078 & 1.059 & 2.279 & 5.994 \\
ECMP & 512 & 0.004 & 0.032 & 0.175 & 0.412 & 0.253 & 0.724 & 1.347 \\
CA & 64 & \textbf{0.000} & 0.001 & 0.013 & 0.050 & 0.020 & 0.092 & 0.311 \\
CA-tuned & 32 & \textbf{0.000} & \textbf{0.000} & 0.005 & 0.031 & 0.013 & 0.059 & 0.208 \\
TLR & $\infty$ & 0.002 & 0.044 & 0.392 & 1.479 & 0.266 & 1.071 & 2.165 \\
DBPR & 32 & 0.002 & 0.028 & 0.151 & 0.473 & 0.328 & 1.130 & 2.082 \\
MA-DRL & $\infty$ & \textbf{0.000}{\scriptsize$\pm$0.000} & 0.005{\scriptsize$\pm$0.001} & 0.040{\scriptsize$\pm$0.004} & 0.106{\scriptsize$\pm$0.004} & 0.083{\scriptsize$\pm$0.012} & 0.254{\scriptsize$\pm$0.021} & 0.619{\scriptsize$\pm$0.070} \\
ELB & 512 & 0.001 & 0.005 & 0.026 & 0.090 & 0.028 & 0.137 & 0.509 \\
\midrule
\textbf{BRACE} & 128 & \textbf{0.000} & \textbf{0.000} & \textbf{0.003} & \textbf{0.017} & \textbf{0.008} & \textbf{0.039} & \textbf{0.150} \\
\midrule
\textit{Centralized MCF}$^\dagger$ & 64 & 0.000 & 0.000 & 0.001 & 0.011 & 0.004 & 0.026 & 0.116 \\
\bottomrule
\end{tabular*}
\begin{tablenotes}[flushleft]
\item[] R and S denote real orbital dynamics and the stress setting, respectively. Numbers indicate the arrival rate in flows/s. Bold marks the best distributed result in each column, including ties. $^{\dagger}$Centralized reference.
\end{tablenotes}
\end{threeparttable}
\end{table*}

\begin{figure*}[!t]
\centering
\includegraphics{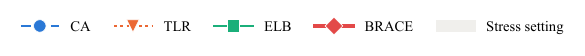}\\[-4pt]
\subfloat[Drop rate]{\includegraphics[width=0.32\textwidth]{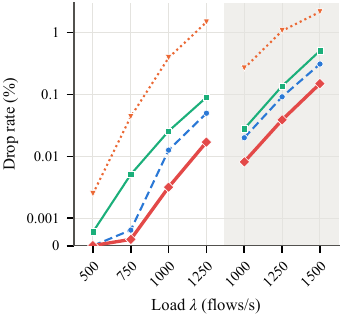}\label{fig:metrics-a}}\hfill
\subfloat[99th-percentile FCT of mice]{\includegraphics[width=0.32\textwidth]{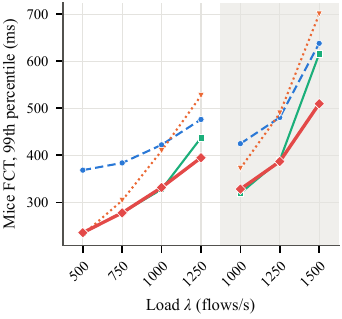}\label{fig:metrics-c}}\hfill
\subfloat[Completion rate of elephants]{\includegraphics[width=0.32\textwidth]{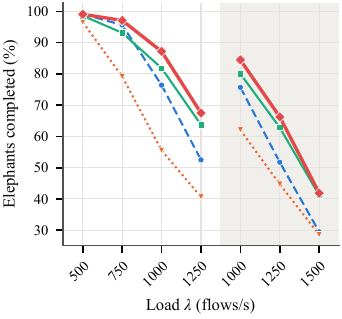}\label{fig:metrics-d}}
\caption{Flow-level metrics with the parameters of Table~\ref{tab:main}.}
\label{fig:metrics}
\end{figure*}

\subsection{Numerical Results}
\label{sec:results}
\subsubsection{Comparison of Slow Layers}
Table~\ref{tab:main} compares the slow layers with their validation-selected parameters.
BRACE obtains the highest reward among the distributed slow layers in all seven conditions and consistently leads ELB by 0.005--0.035.
In the three highest-load conditions, BRACE has a 28--45\% lower drop rate than CA-tuned, the lowest-loss distributed baseline in those conditions.

In these three conditions, BRACE's drop rates are 71--81\% lower than ELB's with the validation-selected parameters and 50--66\% lower under the common $\kappa=8$, where the weight floor is their only difference.
With the validation-selected parameters, BRACE also has a consistently shorter tail FCT of mice than ELB at real load 1250 and stress load 1500, but uses longer paths. At real load 1250, the 99th-percentile FCT is 395 versus 437\,ms and the path stretch is 1.354 versus 1.294. Their mean queueing delays differ by at most 1.01\,ms in all seven conditions.

Fig.~\ref{fig:metrics} shows flow-level results for selected methods. At real loads 500 and 1000 and stress load 1000, BRACE's 99th-percentile FCT of mice is higher than ELB's; it is also higher than that of ECMP, TLR, and DBPR at real load 500 and of CA-tuned at real load 1000. Its mean completion rate for eligible mice is at least 99.9\% in each condition, and it has the highest mean elephant completion rate among the distributed methods except at real load 500, where CA and CA-tuned are slightly higher. At real load 1250, BRACE, ELB, and CA complete 67.5\%, 63.7\%, and 52.5\% of eligible elephants, respectively. BRACE also has the highest credited throughput among the distributed methods at real load 1250 and stress load 1500 (2475 and 2794\,Mbps, respectively).
Across all seven hotspot conditions, BRACE has a lower Jain's fairness index for link utilization than both CA and MA-DRL, indicating less even utilization across links.

With the parameters in Table~\ref{tab:main} held fixed, delaying end-host feedback by one RTT raises BRACE's drop rate to more than ten times its immediate-feedback value in each of the three highest-load conditions. Its drop-rate advantage over CA-tuned narrows from 28--45\% to 13--23\%. Even with this delay, BRACE achieves the highest mean reward among the distributed methods in all seven hotspot conditions.

The validation-selected $(\kappa,\epsilon)$ makes ports with zero raw weight competitive at smaller queue gaps than the common reference (Section~\ref{sec:interface}).
With $\epsilon$ fixed at $10^{-3}$, the same eight-value search selects $\kappa=64$ for BRACE. Its reward lead over validation-selected ELB is consistent only at real loads 500, 750, and 1000.
With the common $\kappa=8$, the mean reward difference between BRACE and ELB ranges from $-0.006$ to $+0.008$, with no consistent lead; in the orbital-geometry control, its absolute value is at most 0.005 across the seven conditions.

The centralized MCF reference has a mean reward that exceeds BRACE's by 0.012--0.254 across the seven hotspot conditions (Table~\ref{tab:main}), and it also has lower path stretch and lower tail FCT of mice in all seven conditions.

\begin{figure*}[!t]
\centering
\includegraphics{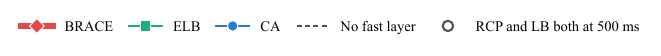}\\[-4pt]
\subfloat[Queue refresh, real load 1250]{\includegraphics[width=0.32\textwidth]{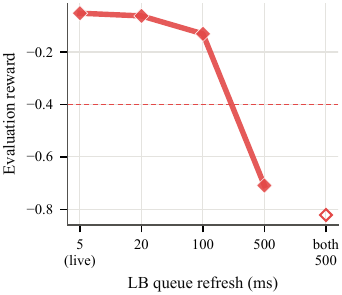}\label{fig:timescale-a}}\hfill
\subfloat[Queue refresh, all loads]{\includegraphics[width=0.32\textwidth]{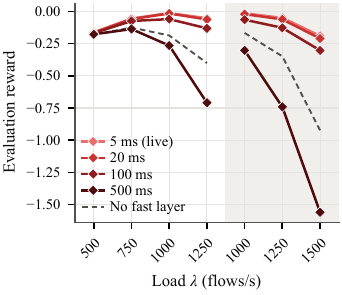}\label{fig:timescale-b}}\hfill
\subfloat[Slow-layer period, real load 1250]{\includegraphics[width=0.32\textwidth]{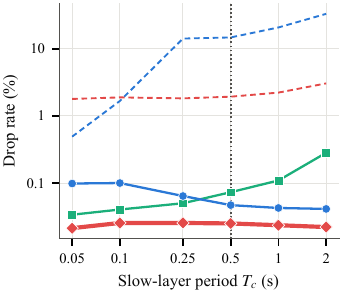}\label{fig:timescale-c}}
\caption{Timescale ablation.
In (a) and (b), the live setting uses current queues at path selection, with a data-plane step of 5\,ms. In (a), ``both 500'' means queue snapshots and RCP rates are updated every 500\,ms. In (c), the dotted line marks the default slow-layer period $T_c$.}
\label{fig:timescale}
\end{figure*}

\begin{figure}[!t]
\centering
\includegraphics{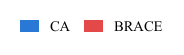}\\[-4pt]
\subfloat[Evaluation reward]{\includegraphics[width=0.49\columnwidth]{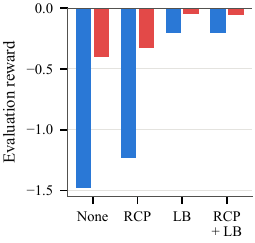}\label{fig:fastlayer-main-reward}}\hfill
\subfloat[99th-percentile FCT of mice]{\includegraphics[width=0.49\columnwidth]{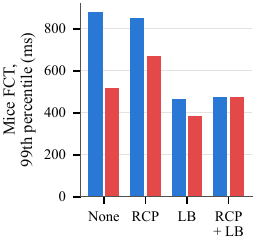}\label{fig:fastlayer-main-fct}}
\caption{Fast-layer ablation for CA and BRACE at real load 1250.}
\label{fig:fastlayer-main}
\end{figure}

\subsubsection{Timescale Separation}
\label{sec:eval-timescale}
Using the common fast-layer reference ($\kappa=8$ for all slow layers, $\epsilon=10^{-3}$ for BRACE), we replace instantaneous queues with snapshots refreshed every 20, 100, or 500\,ms, keeping other settings fixed. A further variant also updates RCP every 500\,ms. Figs.~\ref{fig:timescale}\subref{fig:timescale-a} and~\ref{fig:timescale}\subref{fig:timescale-b} show that a 20-ms snapshot remains close to instantaneous queues, whereas 100\,ms degrades the reward. At real load 1250, BRACE's mean reward is $-0.709$ with 500-ms queue refreshes and $-0.400$ without the fast layer. The same ordering holds at stress load 1500 and, for BRACE at both loads, on all 20 seeds of the orbital-geometry control.

The penalty grows with load: at real load 500, all refresh variants and the case without the fast layer are within 0.018 of each other in reward, whereas 500-ms snapshots are worse than no fast layer on every seed at real loads 1000 and 1250 and all stress loads, for both BRACE and ELB. Stale snapshots repeatedly favor ports that were lightly loaded before the update. At high loads, the fast layer should therefore use queue information refreshed every 20\,ms or faster.

Varying $T_c$ changes both the split-update and access-reassociation intervals (Fig.~\ref{fig:timescale}\subref{fig:timescale-c}). At real load 1250, BRACE's drop rate is 1.79\% at $T_c=50$\,ms without the fast layer, compared with 0.025\% at $T_c=0.5$\,s with the fast layer. The 50-ms setting exchanges neighbor statistics ten times as often. At each tested $T_c$, removing the fast layer increases drop rates on all five test seeds at real loads 1000 and 1250 and stress loads 1250 and 1500. At real load 1250 with the fast layer, BRACE's drop rate remains within 0.021--0.025\% over the tested periods from 0.05 to 2\,s, and ELB's drop rate rises from 0.034\% to 0.283\%.

\subsubsection{Fast-Layer Components}
\label{sec:eval-fast}
Fig.~\ref{fig:fastlayer-main} compares CA and BRACE at real load 1250 under four fast-layer configurations: DCTCP alone (None), RCP, LB, and both. For each matched configuration, BRACE has a higher reward on all five test seeds and a lower seed-averaged 99th-percentile FCT of mice. With both RCP and load balancing, the tail FCT differs by only 0.7\,ms (472.6\,ms for BRACE versus 473.3\,ms for CA), favoring BRACE on four of five seeds. Local load balancing provides most of the gain: at real load 1250, BRACE's reward rises from $-0.400$ without the fast layer to $-0.047$ with load balancing alone. RCP alone reaches $-0.326$. Adding RCP to local load balancing leaves the reward nearly unchanged in the 30-s episodes but lowers losses at the cost of short-flow latency. At real load 1250 and stress load 1500, it reduces drops by 40\% and 52\% and queueing delay by 11\% and 15\%, while raising the 99th-percentile FCT of mice by 22.5\% and 35.6\%. Explicit rates limit new flows from their first sending step, keeping queues shorter while delaying the smallest flows.

\begin{figure}[!t]
\centering
\begingroup
\scriptsize
\definecolor{backlogca}{HTML}{2A78D6}
\definecolor{backlogelb}{HTML}{1BAF7A}
\definecolor{backlogbrace}{HTML}{E34948}
\definecolor{backlogecmp}{HTML}{52514E}
\definecolor{backlogmcf}{HTML}{8C8A85}
\begin{tabular}{@{}lll@{}}
\tikz[baseline=-0.6ex]{\draw[backlogecmp,line width=0.7pt,dash pattern=on 1pt off 1.5pt] (0,0)--(13pt,0);\fill[backlogecmp,shift={(6.5pt,0)}] (90:2.3pt)--(126:0.95pt)--(162:2.3pt)--(198:0.95pt)--(234:2.3pt)--(270:0.95pt)--(306:2.3pt)--(342:0.95pt)--(18:2.3pt)--(54:0.95pt)--cycle;}\,ECMP &
\tikz[baseline=-0.6ex]{\draw[backlogca,line width=0.7pt,dash pattern=on 4pt off 2pt] (0,0)--(13pt,0);\filldraw[fill=backlogca,draw=white,line width=0.3pt] (6.5pt,0) circle (1.7pt);}\,CA &
\tikz[baseline=-0.6ex]{\draw[backlogelb,line width=0.7pt] (0,0)--(13pt,0);\node[fill=backlogelb,draw=white,line width=0.3pt,inner sep=0pt,minimum size=3.2pt] at (6.5pt,0) {};}\,ELB \\
\tikz[baseline=-0.6ex]{\draw[backlogbrace,line width=1.1pt] (0,0)--(13pt,0);\node[fill=backlogbrace,draw=white,line width=0.3pt,inner sep=0pt,minimum size=2.7pt,rotate=45] at (6.5pt,0) {};}\,BRACE &
\tikz[baseline=-0.6ex]{\draw[backlogbrace,line width=0.7pt,dash pattern=on 3pt off 1.2pt] (0,0)--(13pt,0);\node[fill=white,draw=backlogbrace,line width=0.5pt,inner sep=0pt,minimum size=2.5pt,rotate=45] at (6.5pt,0) {};}\,BRACE without fast layer &
\tikz[baseline=-0.6ex]{\draw[backlogmcf,line width=0.7pt,dash pattern=on 4pt off 1.5pt on 1pt off 1.5pt] (0,0)--(13pt,0);\fill[backlogmcf] (5.0pt,-0.5pt) rectangle (8.0pt,0.5pt);\fill[backlogmcf] (6.0pt,-1.5pt) rectangle (7.0pt,1.5pt);}\,Centralized MCF
\end{tabular}\par
\endgroup
\vspace{2pt}
\includegraphics[width=0.83\columnwidth]{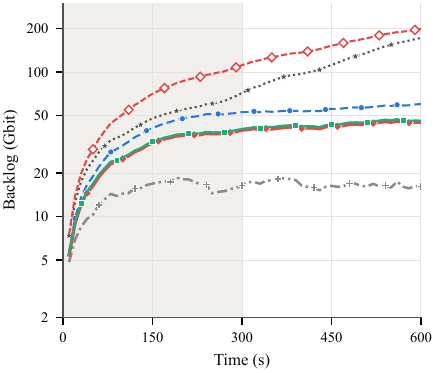}
\caption{Backlog at real load 1250, sampled every 10\,s.
Shading marks the 300-s warm-up.}
\label{fig:long-backlog-main}
\end{figure}

Fig.~\ref{fig:long-backlog-main} shows the buildup of backlog (the total remaining data volume of all active flows) in 600-s episodes at real load 1250 with the common fast-layer reference.
BRACE's mean backlog increases from 12.2\,Gbit at 30\,s to 39.7\,Gbit at 300\,s and 44.9\,Gbit at 600\,s, so the 30-s episodes capture only the early buildup.
Its mean backlog stays below ELB's at every sampled time, while the centralized MCF reference, run on three test seeds (101--103), has a lower backlog.
Over the final 300\,s, adding RCP to local load balancing lowers BRACE's drop rate from 0.093\% to 0.016\% but reduces its reward from $-0.051$ to $-0.278$, because it raises the mean FCT of mice from 193 to 303\,ms.
Whether to enable RCP thus depends on how losses are weighed against short-flow latency.

\begin{figure*}[!t]
\centering
\includegraphics[width=\textwidth]{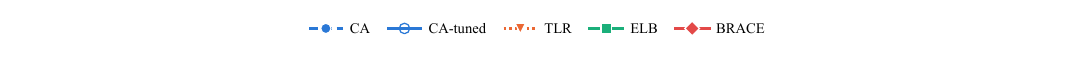}\\[-4pt]
\subfloat[Drop rate]{\includegraphics[width=0.32\textwidth]{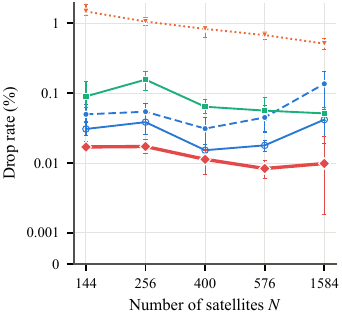}\label{fig:scale-a}}\hfill
\subfloat[Mean FCT of mice]{\includegraphics[width=0.32\textwidth]{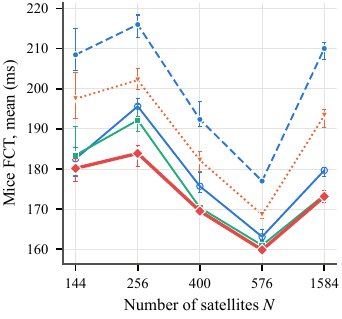}\label{fig:scale-b}}\hfill
\subfloat[99th-percentile FCT of mice]{\includegraphics[width=0.32\textwidth]{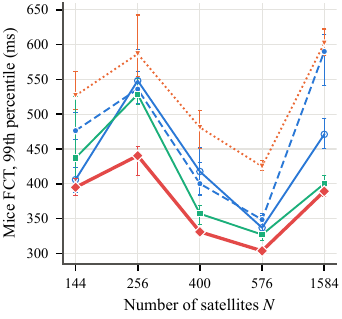}\label{fig:scale-c}}
\caption{Scalability under real orbital dynamics.}
\label{fig:scale}
\end{figure*}

\subsubsection{Scalability}
\label{sec:scalability}

Fig.~\ref{fig:scale} shows selected methods for $N=144$, 256, 400, 576, and 1584 under real orbital dynamics, keeping the inclination, altitude, and traffic model unchanged.
The first four configurations have $12\times12$, $16\times16$, $20\times20$, and $24\times24$ satellites, respectively, in planes of equal size.
The 1584-satellite case uses 72 planes of 22 satellites, following the first Starlink shell geometry~\cite{kassing2020exploring}, with an assumed +Grid ISL topology.
Its shortest paths reach 46 hops, so the data-plane path-length limit is raised from 32 to 64.
Scaling arrivals with $N$ overloads ISLs around hotspot access satellites, and fixed arrivals leave the larger square constellations lightly loaded. We calibrate loads on one validation seed against CA's drop rate at $N=144$ and $\lambda=1250$~flows/s. For $N=144$, 256, 400, 576, and 1584, the resulting loads are 1250, 1500, 1750, 1650, and 1100~flows/s, respectively.
All validated parameters remain fixed, and MA-DRL reuses its network trained at $N=144$.

Across the tested sizes, BRACE's mean reward lead over ELB is 0.014--0.046.
BRACE has a 26--76\% lower drop rate than CA-tuned, the lowest-loss distributed baseline. At $N=1584$, their drop rates are 0.00984\% and 0.04185\%, respectively.
Among the distributed methods, BRACE has the lowest mean and 99th-percentile FCT of mice averaged over seeds at each size. At $N=1584$, its mean FCT is 173.1\,ms versus 173.5\,ms for ELB, and the difference favors BRACE on four of five seeds at both $N=400$ and 1584.
Each satellite computes its own split table. On an Apple M5 Pro, computing all 1584 tables on one thread takes a median of 278\,ms over eight timed control periods after four warm-up periods. Dividing this time by the constellation size gives 0.18\,ms per satellite.

\begin{figure*}[!t]
\centering
\includegraphics[width=\textwidth]{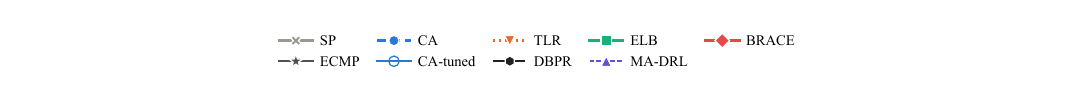}\\[-4pt]
\subfloat[Reward difference to BRACE]{\includegraphics[width=0.49\textwidth]{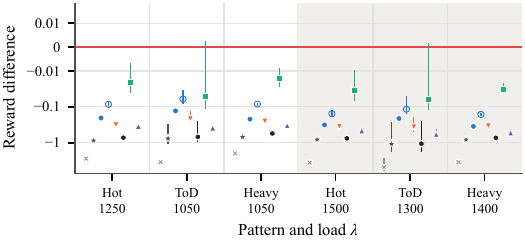}\label{fig:patterns-a}}\hfill
\subfloat[Drop rate]{\includegraphics[width=0.49\textwidth]{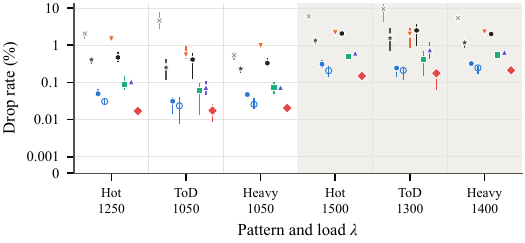}\label{fig:patterns-b}}
\caption{Main hotspot (Hot), time-of-day (ToD), and heavy-tailed (Heavy) traffic patterns. In (a), positive values favor the baseline.}
\label{fig:patterns}
\end{figure*}

\begin{figure*}[!t]
\centering
\subfloat[Validation-selected parameters]{\includegraphics[width=0.49\textwidth]{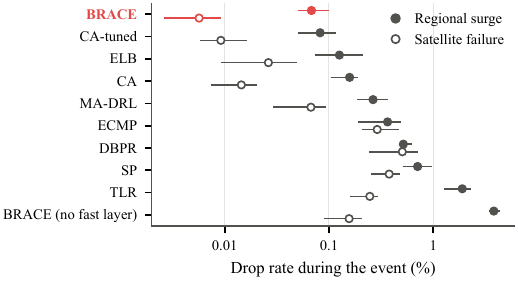}\label{fig:transient-selected}}\hfill
\subfloat[Common fast-layer reference]{\includegraphics[width=0.49\textwidth]{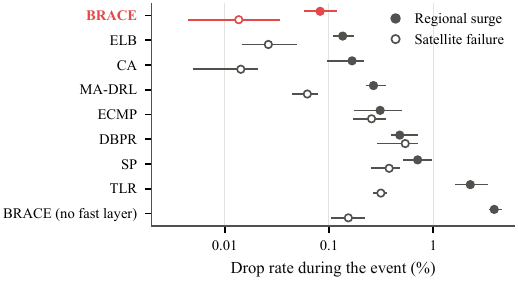}\label{fig:transient-common-main}}
\caption{Drop rate during the regional surge and the satellite failure at real load 1000.}
\label{fig:transient}
\end{figure*}

\subsubsection{Traffic Patterns}
\label{sec:patterns}

On the 144-satellite constellation, we test three alternatives to the main hotspot pattern.
The \emph{time-of-day} pattern multiplies each hotspot weight by $1+0.8\cos(2\pi(h-20)/24)$, where $h$ is local time in hours and each episode's start time in Coordinated Universal Time is drawn from its seed.
The \emph{heavy-tailed} pattern lowers the Pareto shape from 1.2 to 1.05.
The \emph{uniform} pattern draws both endpoints uniformly over the Earth's surface between $\pm55^\circ$ latitude.
The calibration uses CA's mean hotspot drop rates over the three validation seeds at real load 1250 and stress load 1500 as targets. For real dynamics and stress, respectively, the resulting loads are 1050 and 1300 for time-of-day, 1050 and 1400 for heavy-tailed, and 2300 and 2500 for uniform traffic.
Fig.~\ref{fig:patterns} compares reward differences and drop rates for the three hotspot patterns.

With time-of-day and heavy-tailed traffic, BRACE has the highest mean reward and lowest mean drop rate among the evaluated slow layers in all four conditions.
Its reward leads over ELB are consistent for heavy-tailed traffic and positive on four of five seeds for time-of-day traffic. ELB's mean drop rate is 2.4--3.5 times BRACE's.
Under uniform traffic with the hotspot-selected parameters, ECMP consistently leads BRACE in reward by 0.211 under real orbital dynamics and 0.381 under stress.
BRACE's drop rates are 0.0140\% and 0.0571\%, compared with 0.0050\% and 0.0254\% for ECMP. BRACE has 23--25\% higher path stretch.

\subsubsection{Transient Response}
\label{sec:transient}

In separate runs at real load 1000, we inject either a regional surge or a satellite failure during the interval from 10 to 20\,s.
The regional surge doubles arrivals from Tokyo, Shanghai, Beijing, Osaka, Shenzhen--Hong Kong, and Seoul, adding 273~flows/s while leaving the remaining traffic unchanged.
The satellite failure removes all four ISLs of the transit satellite with the highest population-weighted shortest-path betweenness at $t=10$\,s by setting $\zeta_l=1$ (Section~\ref{sec:topology}). We use the same failed satellite for every method.
Fig.~\ref{fig:transient}\subref{fig:transient-selected} reports the drop rate in Eq.~\eqref{eq:drop} over the event window, including discarded queues, for five paired test seeds with the validation-selected parameters held fixed.
During the surge, BRACE loses 0.0679\%, against 0.0823\% for CA-tuned and 0.1259\% for ELB. Without the fast layer, BRACE loses 3.813\%.
Its advantage over CA-tuned is consistent, whereas that over ELB holds on four of five seeds.
BRACE's mean queueing delay rises from 5.5\,ms before the surge to 15.9\,ms during it and averages 5.9\,ms over the 10\,s after it ends.
The failure triggers an average of 29 forced re-routes for BRACE.
BRACE loses 0.0057\%, versus 0.0092\% for CA-tuned and 0.0262\% for ELB. Every other evaluated slow layer consistently loses more.
The mean queueing delays of BRACE, CA, CA-tuned, and ELB do not increase during this failure.

Fig.~\ref{fig:transient}\subref{fig:transient-common-main} repeats the events with a common fast layer ($\kappa=8$, BRACE $\epsilon=10^{-3}$).
During the surge, BRACE has the lowest mean drop rate (0.0823\%) and loses less than every plotted alternative on all five seeds.
During the failure, BRACE also has the lowest mean drop rate (0.01365\%, versus 0.01427\% for CA and 0.02622\% for ELB). Its drop rate is lower than that of CA and ELB on four of the five seeds.

\begin{figure}[!t]
\centering
\begingroup\footnotesize
\definecolor{sensreal}{HTML}{E34948}\definecolor{sensstress}{HTML}{52514E}\definecolor{sensref}{HTML}{52514E}
\tikz[baseline=-0.6ex]{\draw[sensreal,line width=1.1pt] (0,0)--(14pt,0); \filldraw[fill=sensreal,draw=white,line width=0.3pt] (7pt,0) circle (1.7pt);}\,Real 1250\quad
\tikz[baseline=-0.6ex]{\draw[sensstress,line width=1.1pt] (0,0)--(14pt,0); \filldraw[fill=sensstress,draw=white,line width=0.3pt] (7pt,0) circle (1.7pt);}\,Stress 1500\quad
\tikz[baseline=-0.6ex]\draw[sensref,line width=0.8pt,dash pattern=on 0.8pt off 1.2pt] (0,0)--(14pt,0);\,Reference value\par
\endgroup\vspace{1pt}
\subfloat[Weight floor $\epsilon$: reward]{\includegraphics[width=0.49\columnwidth]{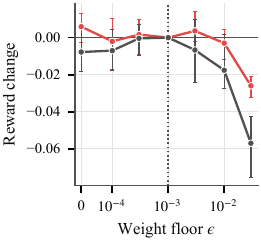}\label{fig:sens-c}}\hfill
\subfloat[Weight floor $\epsilon$: drop rate]{\includegraphics[width=0.49\columnwidth]{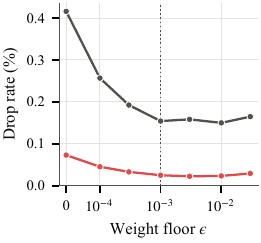}\label{fig:sens-d}}
\caption{Weight-floor sensitivity of BRACE.
(a) Reward change relative to the reference, paired by test seed.}
\label{fig:sens}
\end{figure}

\subsubsection{Parameter Sensitivity}
\label{sec:sensitivity}

We vary one parameter at a time around the common reference ($\kappa=8$, $\epsilon=10^{-3}$) at real load 1250 and stress load 1500.
Rewards are similar for $\epsilon$ between $10^{-4}$ and $10^{-2}$ (Fig.~\ref{fig:sens}). At stress load 1500, the drop rate falls from 0.416\% without the floor to 0.154\% at $\epsilon=10^{-3}$.
At $\epsilon=0.03$, the floor increases detours and lowers the reward. Removing the floor does not change reward consistently at either load.
\looseness=-1

At $\eta=0.9$, the more restrictive rate cap lowers reward by 0.199 at real load 1250 and 0.272 at stress load 1500, and the tail FCT rises above 1.2\,s in both conditions.
Restricting the slow-layer splits to shortest-direction ports, instead of the admissible set in Eq.~\eqref{eq:adm}, lowers reward by 0.705 at real load 1250 and 0.605 at stress load 1500.
\looseness=-1

\section{Conclusion}
\label{sec:conclusion}
Traffic hotspots in LEO networks could overload an inter-satellite link before end-host feedback took effect, while recomputing traffic splits at every queue change would increase onboard computation.
The design challenge was to let each satellite respond to congestion between routing updates using local information while limiting the extra hops incurred by detours.
To address this challenge, we developed BRACE, which periodically assigned positive weights to admissible ports for each destination and used current queue measurements to adjust their selection probabilities in constant time per hop.
In the 144-satellite simulations with immediate end-host feedback, BRACE reduced drop rates by 28--45\% relative to CA-tuned in the three highest-load hotspot conditions.
At high loads, queue measurements refreshed every 20\,ms yielded rewards close to those obtained with instantaneous measurements.
With 500-ms refreshes, stale measurements continued to favor previously underused ports, and mean reward fell below that obtained with the fast layer disabled.

With parameters selected for hotspot traffic, BRACE had higher drop rates and longer paths than ECMP under uniform traffic.
The flow-level model also left packet reordering at flowlet boundaries and forwarding overhead unmeasured.
Future work will test whether adapting the slow-layer thresholds to local queues narrows the gap under uniform traffic.
Packet-level simulations will quantify reordering, measure packet-header and per-flowlet state overhead, and validate the flow-level estimates of short-flow completion time.
Because BRACE uses only link-state information and port occupancies at each satellite and its neighbors, it can be extended to multi-shell constellations with more ISL ports and to space--air--ground integrated networks, where access links can also congest.

\section*{Acknowledgment}
The authors used Claude (Anthropic) and Codex (OpenAI) to improve the grammar, clarity, and readability of the manuscript.
The authors reviewed and verified all revisions made with these tools and take full responsibility for the final content.

\bibliographystyle{IEEEtran}
\bibliography{reference}

%\clearpage
\begin{IEEEbiography}[{\includegraphics[width=1in,height=1.25in,clip,keepaspectratio]{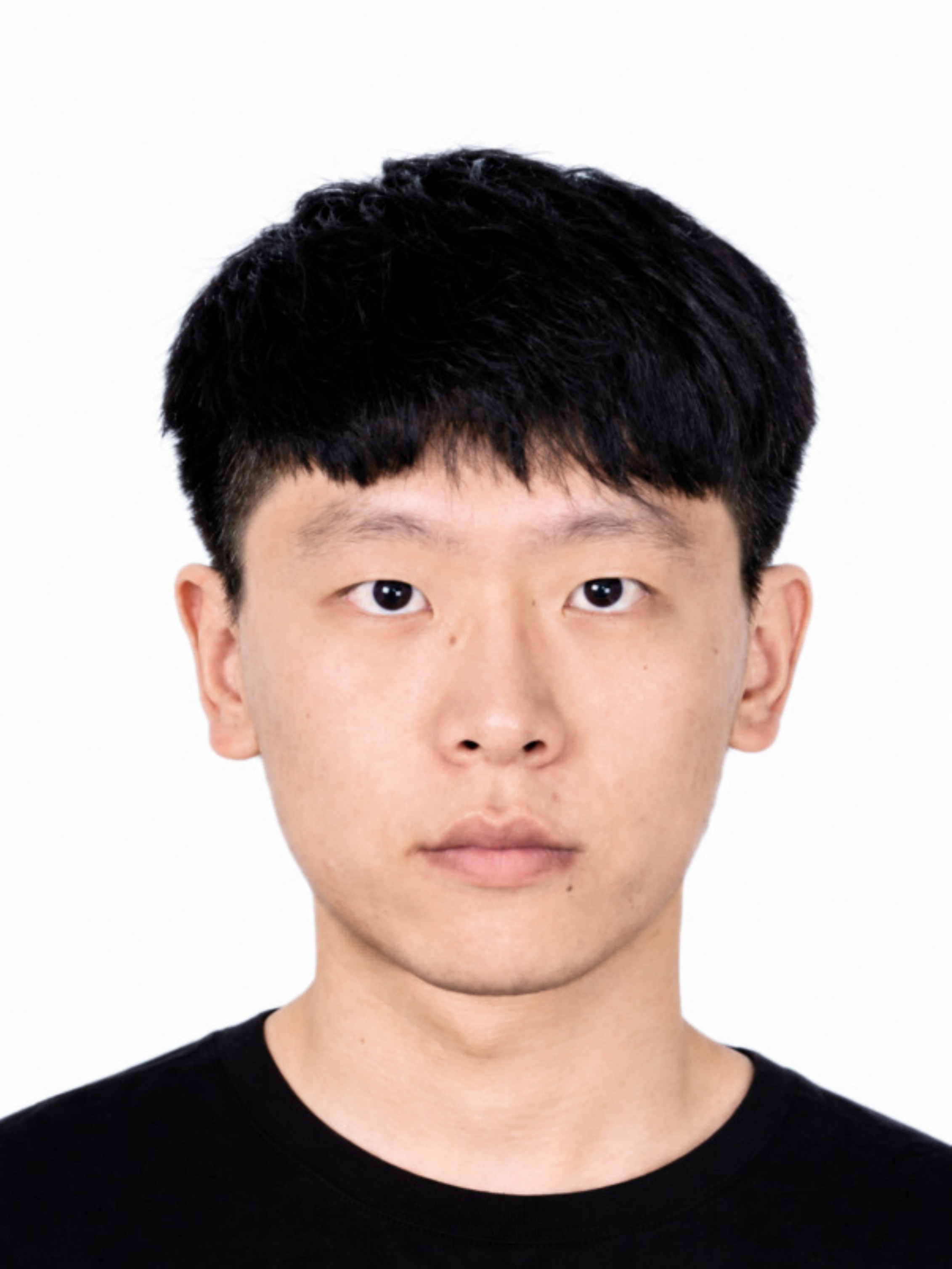}}]{Shuyang Li}
(Member, IEEE) received the Ph.D. degree from Nanjing University of Aeronautics and Astronautics, Nanjing, China, in 2026.
His research interests include non-terrestrial networks, AI-enabled networking, and intelligent optimization.
He received the National Scholarship from the Ministry of Education of the People's Republic of China at the undergraduate, master's, and doctoral levels.
He serves as a reviewer for several leading journals, including IEEE/ACM ToN, IEEE TMC, IEEE TWC, and IEEE TCOM.
\end{IEEEbiography}

\begin{IEEEbiography}[{\includegraphics[width=1in,height=1.25in,clip,keepaspectratio]{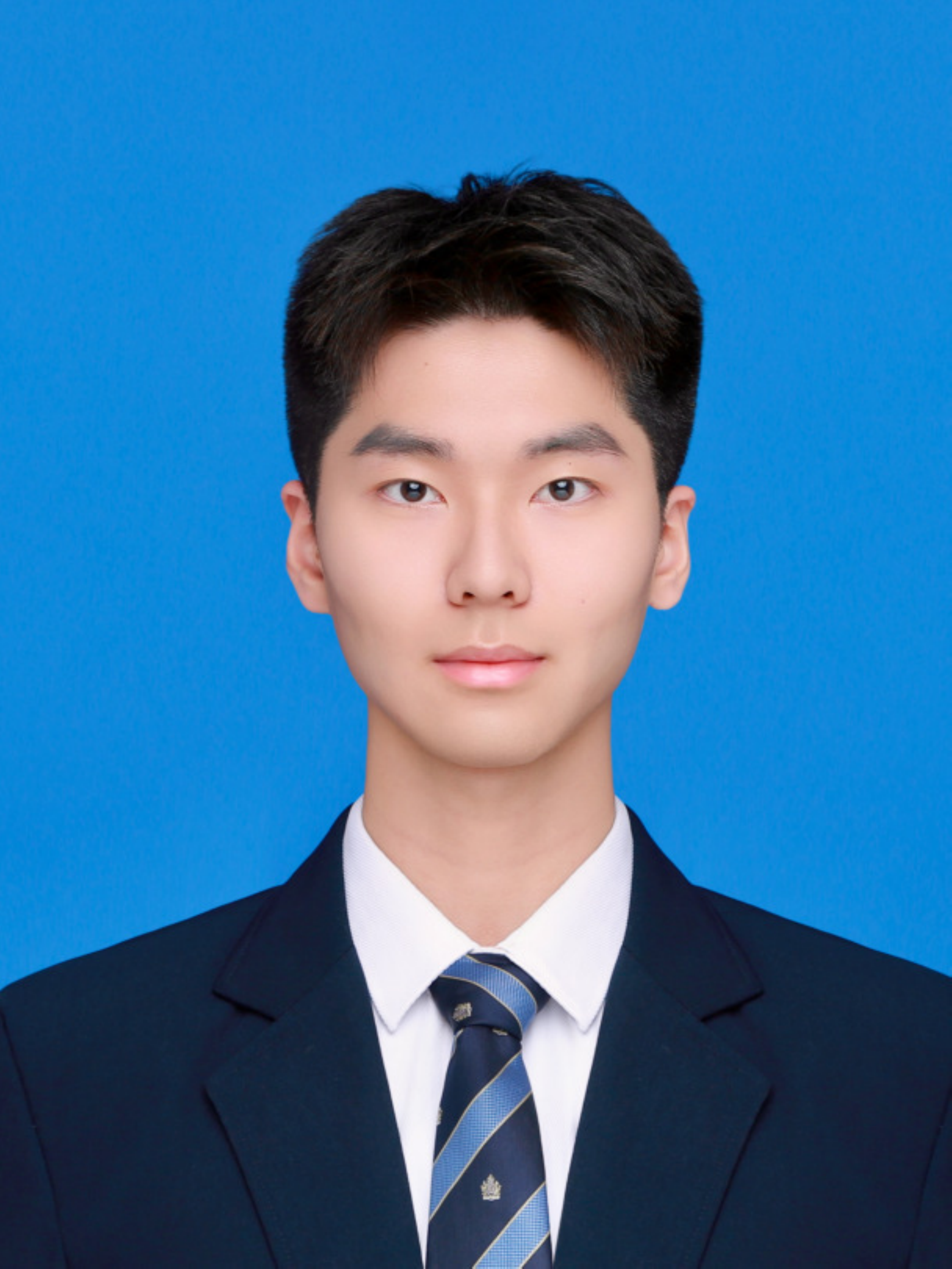}}]{Guanchen Wu}
was born in Nanjing, Jiangsu, China, in 2005.
He is currently pursuing a Bachelor of Science degree at the School of Mathematics, Jilin University, Changchun, China.
His research interests include numerical analysis, mathematical modeling, and multi-objective optimization.
\end{IEEEbiography}

\begin{IEEEbiography}[{\includegraphics[width=1in,height=1.25in,clip,keepaspectratio]{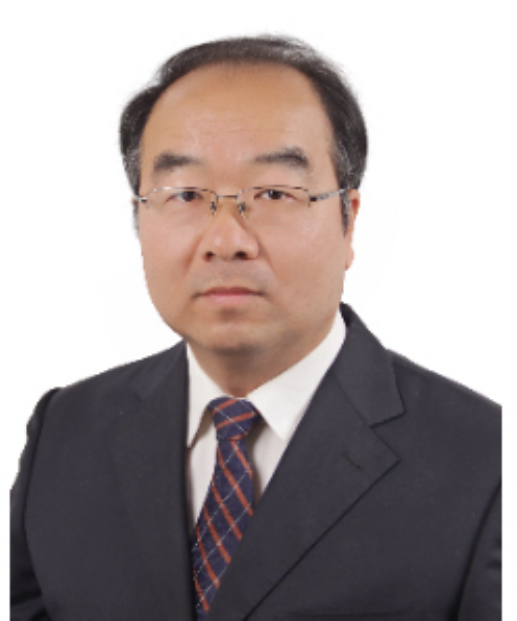}}]{Qiang Wu}
(Member, IEEE) received the Ph.D. degree from the College of Computer Science and Technology, Zhejiang University, Hangzhou, China.
He was a professor with the College of Computer Science and Technology, Nanjing University of Aeronautics and Astronautics, and was also a general engineer with ZTE Central Research Institute.
He is currently a professor with the College of Computer Science and Technology, State Key Laboratory of Blockchain and Data Security, Zhejiang University.
His research interests include future networks, the Industrial Internet, cybersecurity, and space--air--ground integrated networks.
He has been recognized by the Chinese Government with the Second-Class National Science and Technology Progress Award in 2009 and the Second-Class National Technology Innovation Award in 2014.
He holds more than 100 authorized patents, including more than 20 related to international standards.
He is also a fellow of the China Institute of Communications, an executive member of the Service Computing Technical Committee of the China Computer Federation, and a member of the Technical Committee of the National Engineering Laboratory for Mobile Internet System and Application Security.
\end{IEEEbiography}
\vfill

\begin{IEEEbiography}[{\includegraphics[width=1in,height=1.25in,clip,keepaspectratio]{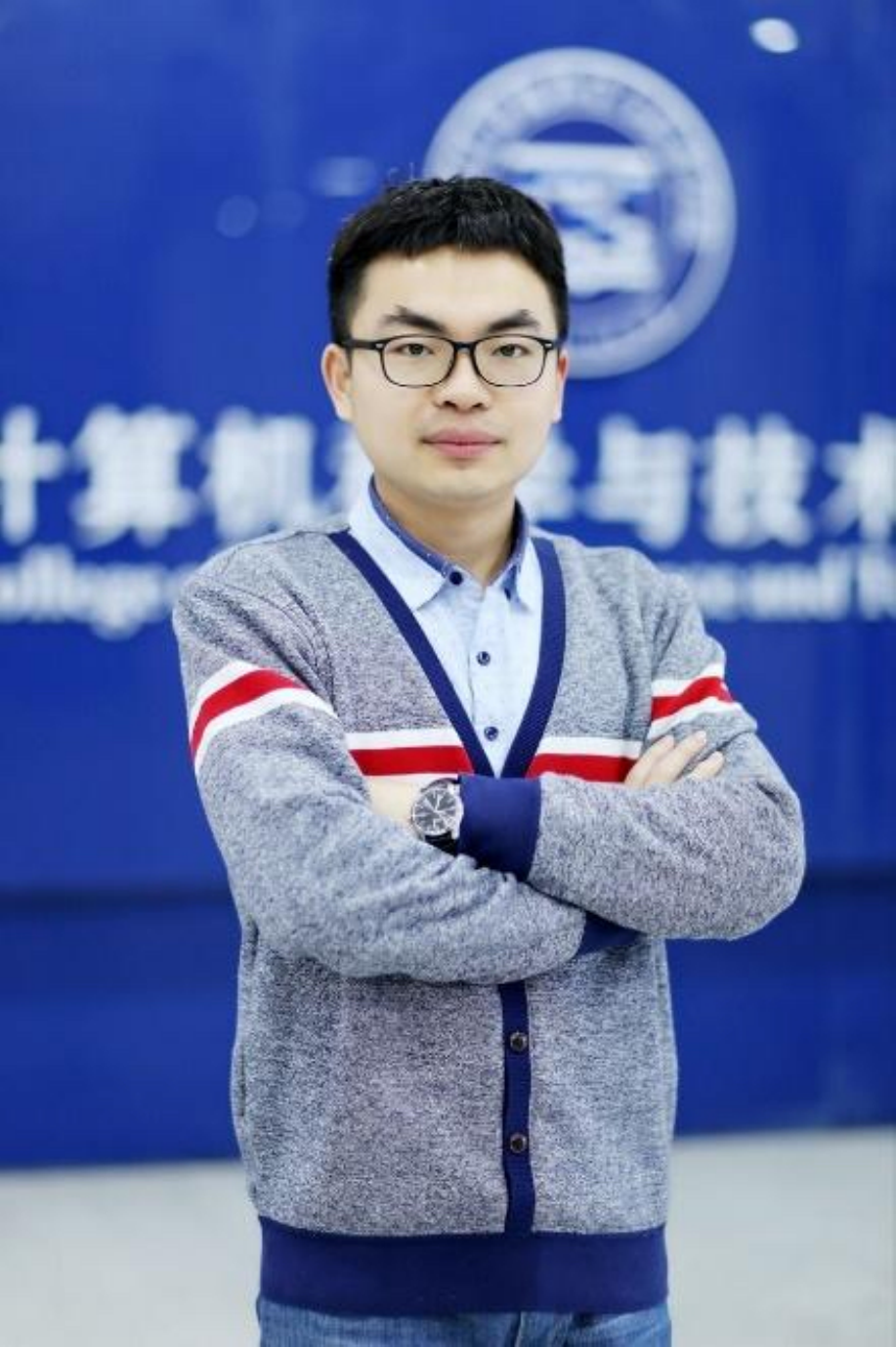}}]{Ran Wang}
(Senior Member, IEEE) is a Full Professor and Doctoral Supervisor with the College of Computer Science and Technology, Nanjing University of Aeronautics and Astronautics, and also with the College of Computer and Data Science, Fuzhou University.
He received a Bachelor of Engineering degree in electronic and information engineering from the Honors School, Harbin Institute of Technology, Harbin, China, in 2011, and the Ph.D. degree in computer science and engineering from Nanyang Technological University, Singapore, in 2016.
He has authored or co-authored over 100 papers in top-tier journals and conferences.
He received the Nanyang Engineering Doctoral Scholarship Award in Singapore and the innovative and entrepreneurial Ph.D. Award of Jiangsu Province, China in 2011 and 2017, respectively.
He received the Second Prize of the Jiangsu Provincial Science and Technology Award (2023) and the Second Prize of the China Communications Society Science and Technology Award (2022).
He received the Best Paper Award at the 2023 ACM Workshop on Mobility in the Evolving Internet Architecture and is a ChangKong Scholar at Nanjing University of Aeronautics and Astronautics.
His current research interests include telecommunication networking and cloud computing.
\end{IEEEbiography}

\begin{IEEEbiography}[{\includegraphics[width=1in,height=1.25in,clip,keepaspectratio]{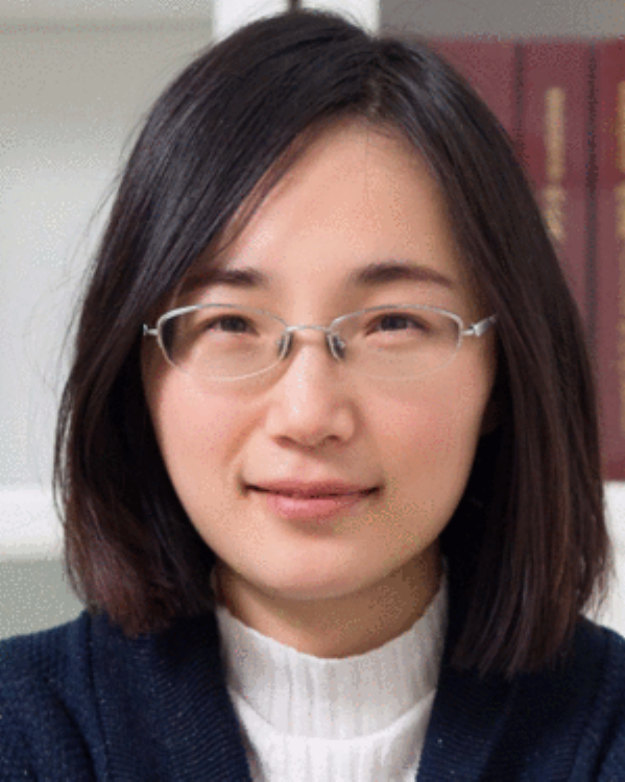}}]{Jie Hao}
(Member, IEEE) received a Bachelor of Science degree from Beijing University of Posts and Telecommunications, Beijing, China, in 2007, and the Ph.D. degree from the University of Chinese Academy of Sciences, Beijing, China, in 2014.
From 2014 to 2015, she was a Post-Doctoral Research Fellow with the School of Computer Engineering, Nanyang Technological University, Singapore.
She is currently an Associate Professor with the College of Computer Science and Technology, Nanjing University of Aeronautics and Astronautics, Nanjing, China.
Her research interests include wireless sensing and wireless communication.
\end{IEEEbiography}

\begin{IEEEbiography}[{\includegraphics[width=1in,height=1.25in,clip,keepaspectratio]{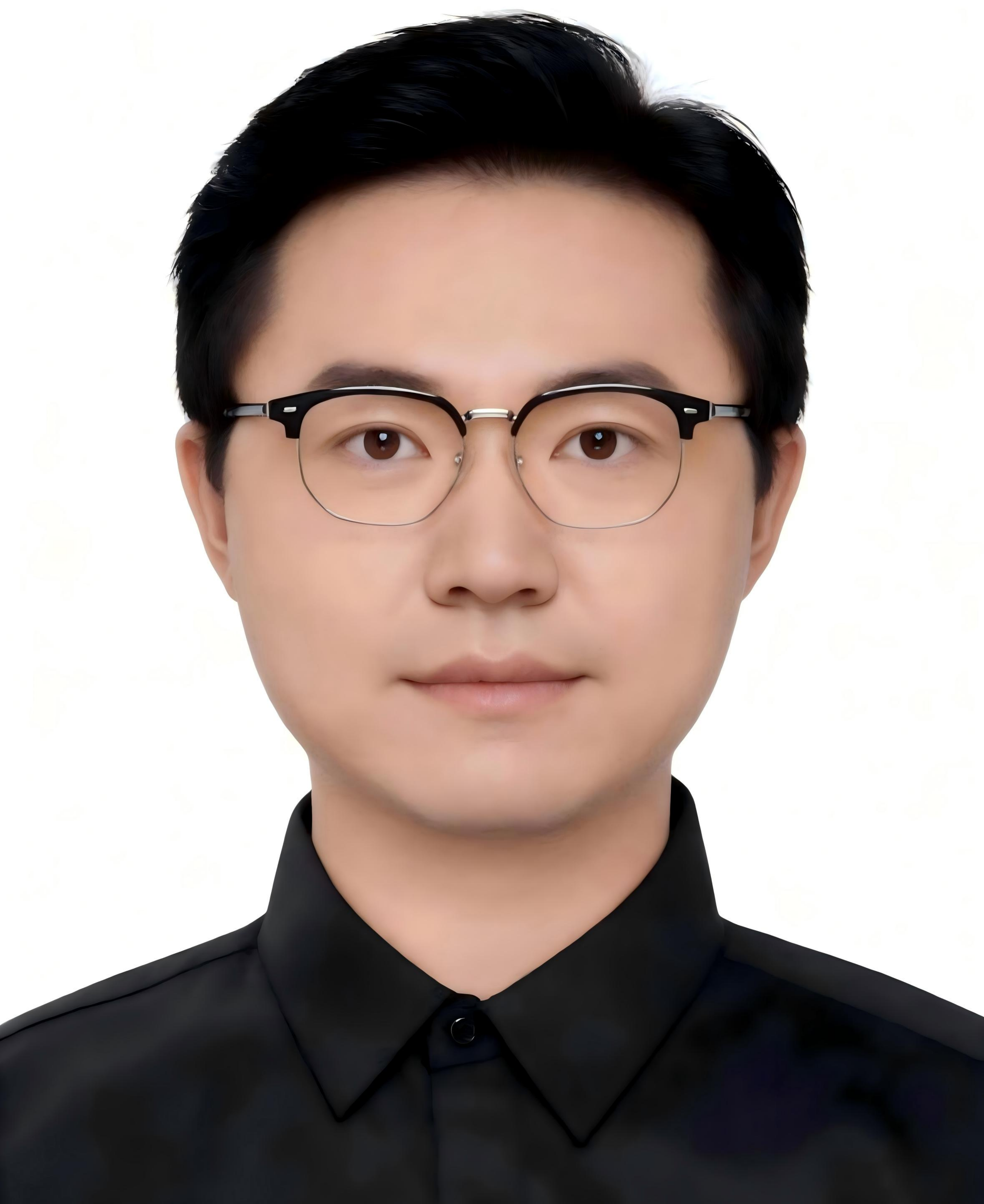}}]{Chao Ren}
(Member, IEEE) received a Bachelor of Engineering degree from Nanjing University of Aeronautics and Astronautics, Nanjing, China, in 2017, and the Ph.D. degree from Nanyang Technological University (NTU), Singapore, in 2022.
He received the Graduate College Research Excellence Award from NTU.
He was a Wallenberg-NTU Presidential Post-Doctoral Fellow with NTU and KTH Royal Institute of Technology.
He is currently with the Aerospace Information Research Institute, Chinese Academy of Sciences, Beijing, China.
His research interests include machine learning, foundation models, quantum computing, and distributed optimization.
\end{IEEEbiography}

\begin{IEEEbiography}[{\includegraphics[width=1in,height=1.25in,clip,keepaspectratio]{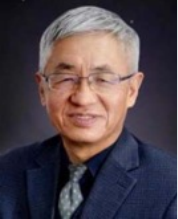}}]{Hongke Zhang}
(Fellow, IEEE) received the Ph.D. degree in communication and information systems from the University of Electronic Science and Technology of China, Chengdu, China, in 1992.
He is currently a Professor in the School of Electronic and Information Engineering at Beijing Jiaotong University, Beijing, China, where he serves as the Director of the National Engineering Center of China on Mobile Specialized Networks.
He is an Academician of the Chinese Academy of Engineering and the Co-Director of the Research Center of Networks and Communications at Peng Cheng Laboratory, Shenzhen, China.
His current research interests include the architecture and protocol design for the future Internet and specialized networks.
Professor Zhang also serves as an Associate Editor for the IEEE Transactions on Network and Service Management and the IEEE Internet of Things Journal.
\end{IEEEbiography}
\vfill

\end{document}